\documentclass[11pt,a4paper]{article}
\usepackage[T1]{fontenc}
\usepackage{authblk}
\usepackage{mathptmx}
\usepackage{stackrel,amssymb,amsmath}
\usepackage{courier}
\usepackage[cm]{fullpage}

\usepackage{cite}

\usepackage{url} 

\usepackage{array}

\usepackage{rotating} 

\usepackage{graphicx}   
\usepackage{longtable}

\usepackage[font=small]{caption}

\usepackage{hyperref} 
\usepackage{xurl}  

\newcommand*{\etal}{\textit{et al.}}

\newcommand{\email}[1]{\texttt{#1}}
\providecommand{\keywords}[1]{{\small\textbf{\textit{Keywords---}} #1}}

\newcommand*{\ksp}{$\mathrm{K\text{-}S_P}$}
\newcommand*{\ksa}{$\mathrm{K\text{-}S_A}$}
\newcommand*{\kcp}{$\mathrm{K\text{-}C_P}$}
\newcommand*{\kca}{$\mathrm{K\text{-}C_A}$}
\newcommand*{\card}[1]{\lvert #1 \rvert}

\DeclareMathOperator{\graphdeg}{deg}

\newcommand{\astfootnote}[1]{%
\let\oldthefootnote=\thefootnote%
\setcounter{footnote}{0}%
\renewcommand{\thefootnote}{\fnsymbol{footnote}}%
\footnote{#1}%
\let\thefootnote=\oldthefootnote%
}

\begin{document}

\emergencystretch 3em

\title{Improving exponential-family random graph models for bipartite networks}

\author[1,*]{Alex Stivala}
\author[2,1]{Peng Wang}
\author[3]{Alessandro Lomi}

\affil[1]{Institute of Computing, Universit\`a della  Svizzera italiana, Via Giuseppe Buffi 13, 6900 Lugano, Switzerland}
\affil[2]{Centre for Transformative Innovation, Swinburne University of Technology, John Street, Hawthorn VIC 3122, Australia}
\affil[3]{Universit\`a della  Svizzera italiana, Via Giuseppe Buffi 13, 6900 Lugano, Switzerland}
\affil[*]{Corresponding author: \email{alexander.stivala@usi.ch}}

\maketitle

\begin{abstract}
{Bipartite graphs, representing two-mode networks, arise in many research fields. These networks have two disjoint node sets representing distinct entity types, for example persons and groups, with edges representing associations between the two entity types. In bipartite graphs, the smallest possible cycle is a cycle of length four, and hence four-cycles are the smallest structure to model closure in such networks. Exponential-family random graph models (ERGMs) are a widely used model for social, and other, networks, including specifically bipartite networks. Existing ERGM terms to model four-cycles in bipartite networks, however, are relatively rarely used. In this work we demonstrate some problems with these existing terms to model four-cycles, and define new ERGM terms to help overcome these problems. The position of the new terms in the ERGM dependence hierarchy, and their interpretation, is discussed. The new terms are demonstrated in simulation experiments, and their application illustrated on a canonical example of an empirical two-mode network.}
\end{abstract}


\section{Introduction}
\label{sec:intro}

Bipartite graphs are graphs whose nodes can be partitioned into two
disjoint sets, such that an edge exists only between nodes in
different sets. Such graphs have important applications in
representing two-mode networks, which are networks in which there are
two types of nodes, with edges possible only between nodes of
different types. An important example of a two-mode network is an
affiliation network, in which one type of node represents a person, the
other type of node represents a group, and an edge represents
membership of a person in a group \cite{breiger74}. Two-mode networks have applications not only in sociology, but
also biology, ecology, political science, psychology, finance, and economics; for a
recent review of applications and methods for two-mode networks, see
Neal \etal{}~\cite{neal24c}.
Bipartite networks also arise as representing the meso-level network
in the conceptualization and analysis of multilevel networks
\cite{wang13_multilevel}.

Two-mode networks can be studied by means of their projections onto
one-mode (unipartite) networks, thereby allowing the use of existing
methods for one-mode networks, however this can result
in lost information, and properties of the one-mode networks (such as
high clustering coefficients) that are due to the projection process
rather than the original data \cite{latapy08}.  Although the former
problem can be ameliorated by using both projections in analyses
\cite{everett13,everett16}, it is still desirable to study
the original two-mode network directly, for which specific methods are required
\cite{latapy08}.

In studying one-mode networks, a central concept is triadic closure,
the tendency for a path of length two (a ``two-path''; three nodes
connected by two edges) to be ``closed'' into a triangle by the
addition of a third edge. In the context of social networks, this is
the process of a friend of a friend becoming themselves a friend, and is
perhaps most well known via the ``strength
of weak ties'' \cite{granovetter73} argument, whereby an open two-path
of strong ties is the ``forbidden triad'', which is ``forbidden''
because the two actors with strong ties to a common third actor must
themselves have a strong tie.
In a bipartite network, however, a closed triad (triangle) is impossible;
indeed it is a defining feature of bipartite graphs that only cycles of even length are possible \cite{mathworld_bipartite_graph}. Therefore, the
smallest possible cycle in a bipartite graph is a four-cycle, and hence
four-cycles are frequently used to measure closure in bipartite
networks \cite{robins04b,opsahl13}.

An example of a social process that creates four-cycles is peer
referral in director interlock networks
\cite{robins04b,koskinen12}. If a director on the boards of two
companies recruits a director from one of them to also sit on the
board of the other, then an open three-path is closed, forming a
four-cycle.

Four-cycles in bipartite networks also have
the particular importance that, together with the degree distribution,
they explain the degree assortativity in the one-mode projected
network \cite{vasquesfilho20,vasquesfilho20b}.

Exponential random graph models (ERGMs) are a widely used model for
social \cite{lusher13,amati18,koskinen20,koskinen23}, and other
\cite{cimini19,ghafouri20,giacomarra23}
networks. Specific forms of the ERGM have been developed for two-mode
networks
\cite{wang09,wang13_bipartite,wang13_in_lusher13book,bomiriya23},
however, as we shall show in this work, existing ERG models for
bipartite networks often have problems modelling four-cycles, and
hence can frequently not adequately model closure in bipartite networks.

In this work, we will show that, despite the importance of four-cycles in two-mode networks, ERGM terms to model four-cycles in such networks are
relatively rarely used, in contrast to the ubiquity of terms modelling triadic closure in one-mode networks. We will then describe some problems with existing configurations for modelling four-cycles that could explain
their relatively infrequent use, and propose new ERGM configurations for modelling
four-cycles to help overcome these problems. We will discuss the
interpretation of these new parameters, and their position in the
dependence hierarchy of Pattison \& Snijders
\cite{pattison13_in_lusher13book}. We will then demonstrate the new
configurations using simulation experiments and demonstrate their use on
a canonical example of an empirical two-mode network.

\section{The exponential-family random graph model (ERGM)}
\label{sec:ergm}

An ERGM is a probability distribution with the form
\begin{equation}
  \label{eqn:ergm}
  \Pr_\theta(X = x) = \frac{1}{\kappa(\theta)}\exp\left(\sum_C \theta_C g_C(x)\right)
\end{equation}
where
\begin{itemize}
\item $X = [X_{ij}]$ is square binary matrix of random tie variables,
\item $x$ is a realization of $X$,
\item $C$ is a ``configuration'', a set of nodes and a subset of ties between them, designed in order to model a particular structure of interest,
\item $g_C(x)$ is the network statistic for configuration $C$,
\item $\theta$ is a vector of model parameters, where each $\theta_C$ is the parameter corresponding to configuration $C$,
\item $\kappa(\theta) = \sum_{x \in G_n} \exp\left(\sum_C \theta_C
  g_C(x)\right)$, where $G_n$ is the set of all square binary matrices
  of order $n$ (graphs with $n$ nodes), is the normalising constant to
  ensure a proper distribution.
\end{itemize}

We will use the notation $x_{ij}$ (where $1 \leq i \leq n$, $1 \leq j \leq n$) for elements of the binary adjacency matrix $x$. In this work we consider only the case of undirected networks, so $x_{ij} = x_{ji}$, and the cardinality of $G_n$ is $\card{G_n} = 2^{n(n-1)/2}$. In the case of bipartite networks, the two disjoint node sets are denoted $A$ and $B$, with sizes $N_A = \card{A}$ and $N_B = \card{B}$ respectively, and so $N_A + N_B = n$, and $x_{ij} = 0$ if both $i$ and $j$ are in node set $A$ or both $i$ and $j$ are in node set $B$. In this bipartite case, the normalising constant is
$\kappa(\theta) = \sum_{x \in G^{\mathrm{Bipartite}}_{N_A,N_B}} \exp\left(\sum_C \theta_C g_C(x)\right)$, where $G^{\mathrm{Bipartite}}_{N_A,N_B}$ is the set of all bipartite graphs with node set sizes $N_A$ and $N_B$. This set has cardinality $\card{G^{\mathrm{Bipartite}}_{N_A,N_B}} = 2^{N_A N_B}$.

Estimating the value of the parameter vector $\theta$ which maximises the probability of the observed graph, that is, the maximum likelihood estimator (MLE), enables inferences regarding the under-representation (negative and statistically significant estimate) or over-representation (positive and statistically significant estimate) of the corresponding configurations. These inferences are conditional on the other configurations included in the model, which need not be independent.

Estimating the MLE of (\ref{eqn:ergm}) is computationally intractable due to the normalising constant $\kappa(\theta)$ (specifically, the size of the set of graphs it sums over). Therefore, Markov chain Monte Carlo (MCMC) methods are usually used \cite{geyer92,snijders02,hunter12}. One such algorithm is the ``Equilibrium Expectation'' (EE) algorithm \cite{byshkin16,byshkin18,borisenko20}, which was recently shown to converge to the MLE \cite{borisenko20,giacomarra23}.

ERG models with only simple configurations (such as the Markov edge plus triangle model) can be prone to problems with phase transitions or ``near-degeneracy'' \cite{handcock03,snijders06,fellows17,schweinberger11,chatterjee13,schweinberger20,blackburn23,koskinen23}. Such problems are typically avoided by the use of ``alternating'' \cite{snijders06,robins07,koskinen13} or ``geometrically weighted'' \cite{hunter07,stivala23_alaam} configurations. The former are parameterised with a decay parameter $\lambda$ controlling the rate at which the weight of contributions from additional terms in the statistic decay. The corresponding parameter for the geometrically weighed configurations can be estimated as part of the model, in which case it becomes a ``curved ERGM'' \cite{hunter06}, however in this work we will use fixed values of $\lambda$ for the ``alternating'' configurations.

In order to model two-mode networks, and in particular affiliation networks, Wang \etal~\cite{wang09} define the alternating $k$-two-path
statistics \kca{} and \kcp{}, implemented as XACA and XACB in MPNet
\cite{mpnet14,mpnet22}, and BipartiteAltKCyclesA and
BipartiteAltKCyclesB in EstimNetDirected
(\url{https://github.com/stivalaa/EstimNetDirected}):
\begin{equation}
  \label{qqn:bipartitealtkcyclesa}
  z_{\mathrm{K\text{-}C_A}(\lambda)} = z_{\mathrm{XACA}(\lambda)} = z_{\text{BipartiteAltKCyclesA}(\lambda)} = \lambda \sum_{i \in B} \sum_{\left\{l \in B\,:\,l<i\right\}} \left[ 1 - \left( 1 - \frac{1}{\lambda} \right)^{L_2(i,l)} \right]
\end{equation}
where $\lambda > 1$ is the decay parameter, and $L_2(i,l)$ is the number of two-paths connecting $i$ to $l$:
\begin{equation}
  \label{eqn:L2}
  L_2(i,l) = \sum_{h \neq i,l} x_{ih} x_{hl}.
\end{equation}
The corresponding change statistic is \cite{wang09}:
\begin{align}
  \label{eqn:change_bipartitealtkcyclesa}
  \delta_{\mathrm{K\text{-}C_A}(\lambda)}(i,j) = \delta_{\mathrm{XACA}(\lambda)}(i,j) =   \delta_{\text{BipartiteAltKCyclesA}(\lambda)}(i,j) &= \sum_{l \in B} \left[ x_{il} \left( 1 - \frac{1}{\lambda}\right)^{L_2(j,l)} \right] \\
  &= \sum_{l \in N(i)} \left( 1 - \frac{1}{\lambda}\right)^{L_2(j,l)}.
\end{align}
where $N(i)$ denotes the neighbours of node $i$, that is, nodes
$k \neq i$ such that $x_{ik} = 1$, or, equivalently, $d(i,k) = 1$,
where $d(u,v)$ is the geodesic distance from $u$ to $v$.
The statistic and change statistic for BipartiteAltKCyclesB (\kcp{} or XACB) are defined similarly.

\section{Literature survey}
\label{sec:litreview}

In order to get an overview of which effects are used in modelling bipartite networks with ERGMs, and how well they fit four-cycles, we conducted a comprehensive survey of publications which included ERG models of bipartite networks.
To be included, a publication must contain one or more ERG models of one or more empirical bipartite (two-mode) networks. There must be sufficient detail given to know at least the parameters included in the models, and their estimated signs and statistical significance. Models of one-mode projections of two-mode networks were excluded; we only consider ERG models of the bipartite network itself. Models of multilevel networks were excluded, although if there is a model of just the cross-level (bipartite) network, this is included.
The papers included in the survey are listed in Table~\ref{tab:literature} in Appendix~\ref{sec:survey_models}, which also contains further details
of the literature survey.

Most of the models in Table~\ref{tab:literature} were estimated with the BPNet \cite{bpnet}, MPNet  \cite{mpnet14,mpnet22} or statnet \cite{handcock08,hunter2008ergm,hummel12,statnet,ergm,ergm4,ergm4_computational} software, but a small number were estimated either by maximum pseudo-likelihood estimation (MPLE) or with Bayesian methods using the Bergm \cite{caimo14,bergm} software. The one model included that was estimated with Bergm contains no terms to model four-cycles or goodness-of-fit tests including four-cycles \cite{balest19}. Of the ten models (across three publications) estimated by MPLE, only one contains a term to model four-cycles, and this is found to be positive and significant \cite{agneessens08}.

Table~\ref{tab:literature_number_estimated_bpnet} summarises the parameter
estimates in models from the literature in Table~\ref{tab:literature} that were estimated using BPNet or MPNet, and contain the four-cycles parameter C4, or the bipartite
alternating $k$-two-path parameters (\kcp{}  and  \kca{}) defined in Wang \etal{}~\cite{wang09}. Less than a third
($20/63$) include the four-cycles parameter, less than half ($30/63$)	
include either of the two alternating $k$-two-path parameters,
and less than 30\% ($17/63$) include both 
\kcp{} and \kca{}.

\begin{table}[!htbp]
  \caption{Counts of parameters in models estimated by BPNet or MPNet (total 63) in the reviewed literature.}
    \label{tab:literature_number_estimated_bpnet}
    {\begin{tabular*}{\textwidth}{@{\extracolsep{\fill}}lrrr@{}}
         \hline
         & C4 & \kcp{} & \kca{} \\
         \hline
         Total estimated & 20    & 30    & 25   \\
         Negative & 14    & 20    & 15   \\
         Negative and signif. & 5     & 14    & 9 \\
         Positive & 6     & 10    & 10    \\
         Positive and signif. & 4     & 7     & 2 \\
         \hline
     \end{tabular*}}{}
\end{table}

Table~\ref{tab:literature_number_estimated_statnet} summarises the
parameter estimates in models from the literature in
Table~\ref{tab:literature} that were estimated using statnet, and
which contain the four-cycle term, the bipartite geometrically
weighted dyadwise shared partner distribution (gwb1dsp or gwb2dsp)
term, the statnet equivalent of the \kcp{} and
\kca{} parameters, or the geometrically weighted
non-edgewise shared partner (gwnsp) term.
Only one model estimated with statnet contained an
explicit term for four-cycles \cite{benton17}. Further, 
only $5/43$ models contain a gwb1dsp pr gwb2dsp
term at all, and only two models contain both (Lubell \& Robbins \cite{lubell22}
have two models, both of which include both gwb1nsp and gwb2nsp).

\begin{table}[!htbp]
  \caption{Counts of parameters in models estimated by statnet (total 43) in the reviewed literature. The counts for gwb1dsp and gwb2dsp include those for the equivalent parameters gwb1nsp and gwb2nsp, respectively.}  
    \label{tab:literature_number_estimated_statnet}
    {\begin{tabular*}{\textwidth}{@{\extracolsep{\fill}}lrrrr@{}}
         \hline
                              & cycle(4) & gwb1dsp & gwb2dsp & gwnsp   \\
         \hline
         Total estimated &       1 & 2     & 5     & 4   \\
         Negative &              1 & 1     & 2     & 1   \\
         Negative and signif. &  1 & 1     & 0     & 0   \\
         Positive &              0 & 1     & 3     & 3   \\
         Positive and signif. &  0 & 1     & 2     & 3   \\
         \hline
     \end{tabular*}}{}
\end{table}

Less than a quarter ($26/117$) of the models include an explicit assessment of goodness-of-fit
to four-cycles, and of those,  the majority ($18/26$) are good. Only seven of these are for models that explicitly include C4 as a model parameter, and, as expected of any converged model containing this term, these models fit four-cycles well. Conversely, all of the models which are described as having a poor fit to four-cycles in the goodness-of-fit procedure are models that do not contain the C4 parameter. However, of these eight models, five contain either \kca{} or \kcp{}, and one contains both.

This relative rarity of models containing terms to model closure (four-cycles) in bipartite networks, or assess goodness-of-fit to four-cycles, is in stark contrast to ERG modelling for one-mode networks, where terms modelling triadic closure, such as triangles, alternating-$k$-triangles, or geometrically weighted edgewise shared partners (gwesp in statnet) are almost always included in models, since triadic closure (as evidenced by clustering, or transitivity, in the network), is a well-known feature of social networks \cite{granovetter73,watts98,newman03c}. For example, Clark \& Handcock \cite{clark22} use the latent order logistic (LOLOG) model \cite{fellows18} to reproduce ERG models for 13 networks from peer reviewed papers, and all of these models contain alternating-$k$-triangles \cite{robins07,pauksztat11,doreian12,wong15}, gwesp \cite{goodreau07,heidler14,sailer12,fischer15,toivonen09,ackland111}, or three-cycles \cite{pauksztat11,anderson99b} terms, and in the majority of cases the results are able to be replicated with a triangle term in the LOLOG model \cite{clark22}. And yet, despite higher than expected numbers of four-cycles (or bipartite clustering coefficient) being a notable feature of some bipartite networks, such as director interlock networks \cite{robins04b,opsahl13}, and collaboration and communication networks \cite{opsahl13}, the literature survey presented here shows that terms to model four-cycles are relatively rarely used in published ERG models of bipartite networks.
This could be due to researchers choosing not to model bipartite closure,
however, as shown by some examples in the following section, it can also
be due to difficulties in obtaining converged model estimates when using
existing model terms.

\section{Problems with existing bipartite ERGM statistics}
\label{sec:problems}

It is notable that in the original paper proposing \kca{} and \kcp{} configurations \cite{wang09}, they are explicitly described as $k$-two-path statistics, and in fact they are just bipartite versions of the one-mode alternating-$k$-two-path statistic, representing multiple shared partners (gwdsp in statnet, and then later gwb1dsp and gwb2dsp for bipartite networks). There is, however, already some ``semantic slippage'' into interpreting them as ``cycles'' or ``closure'' --- even though they only actually include cycles (closure) when $k > 1$ ($k = 1$ is a two-path, $k = 2$ is a four-cycle; see Fig.~3 in Wang \etal{}~\cite{wang09}). For example, ``... a better chance of achieving model convergence when closure effects ($K-C_P$ and $K-C_A$) are included in the model.'' \cite[p.~22]{wang09}. Even the name \kca{} (or \kcp{}) suggests ``cycle'' or ``closure'' by the use of the ``C'' (C4 is used for four-cycles in the paper).
Wang \etal{}~\cite{wang13_bipartite} goes back to naming the \kcp{} and \kca{} statistics as A2P-A and A2P-B and describing them as ``shared affiliations (alternating two-paths)'' \cite[p.~213]{wang13_bipartite}. However the MPNet terminology is XACA and XACB \cite{mpnet14,mpnet22}, again with the ``C'' suggestive of cycles or closure, and the EstimNetDirected \cite{stivala20} software (\url{https://github.com/stivalaa/EstimNetDirected}) refers to these effects as BipartiteAltKCyclesA and BipartiteAltKCyclesB.

In the influential book edited by Lusher \etal{}~\cite{lusher13}, Wang \cite{wang13_in_lusher13book} describes \kca{}  and \kcp{} explicitly as ``alternating A cycles'' and ``alternating P cycles'' \cite[p.~124]{wang13_in_lusher13book}, with \cite[Fig.~10.11]{wang13_in_lusher13book} captioned ``Alternating 2-paths'' but with the figure panels labelled ``A cycles (KCA)'' and ``P cycles (KCP)'' respectively \cite[p.~125]{wang13_in_lusher13book}. This description or interpretation is carried over into the empirical part of the book, with Harrigan \& Bond \cite{harrigan13_in_lusher13book}, applying bipartite ERGM to a director interlock network, describing \kcp{} and \kca{} as ``alternating $k$-cycles'' for directors and corporations respectively \cite[p.~266]{harrigan13_in_lusher13book}, and in the ERGM results tables as ``Director 4-cycles'' and ``Corporation 4-cycles'' \cite[pp.~268-269]{harrigan13_in_lusher13book}.
It is also notable that Harrigan \& Bond \cite{harrigan13_in_lusher13book} describes the difficulty of fitting models with \kca{} and \kcp{} and the resulting poor goodness-of-fit for the four-cycle statistic:

\begin{quote}
We found for this particular network that we cannot have both K-Cp and K-Ca in the same model due to convergence issues. In line with Wang, Sharpe, Robins, and Pattison (2009), we present two alternative models, one with each possible k-cycle parameter.
...
However, in line with Wang, Sharpe, Robins, and Pattison (2009), we had a poor fit on the classic 4-cycle parameter [C4], which suggests that improving these structural effects is a substantial area of future research.
\cite[p.~267]{harrigan13_in_lusher13book}
\end{quote}

Given the results of the literature review described in Section~\ref{sec:litreview} it would seem, however, that no such research improving these structural effects has been published yet, and this work may be the first attempt to do so.

We may conclude from this that the \kca{} and \kcp{} statistics count too many things other than four-cycles to be usefully used and interpreted as bipartite closure in many cases. Specifically, they count simple two-paths ($k$-two-paths with $k=1$) as their first (highest weighted) term. This results in situations where long paths or cycles contribute to the \kcp{} and \kca{} statistics, despite having exactly zero four-cycles (see Table~\ref{tab:examples}). Perhaps even more problematically, high-degree nodes, or stars, for example the ``Nine-star'' structure in Table~\ref{tab:examples}, result in large values of \kca{} or \kcp{} (XACA or XACB in MPNet terminology), depending on which node set the hub node is in, but also have exactly zero four-cycles. It seems clear that large networks are likely to contain many stars, and many paths (of length two or more; and note that a two-star is just a two-path), as well as large cycles \cite{stivala20c,stivala23_geodesic}, and these will contribute to large values of the \kcp{} and/or \kca{} statistics, but nothing to the number of four-cycles (the C4 statistic). 

\def \examplefigscale {0.09} 
\begin{table}[htbp!]
  \centering
  \caption{Statistics of some example bipartite networks. L is the number of edges, C4 the number of four-cycles, and BpNP4CA and BpNP4CB are the new statistics BipartiteFourCyclesNodePowerA and BipartiteFourCyclesNodePowerB. Nodes in node set A are represented as red circles, and nodes in node set B as blue squares.}
    \begin{tabular}{lm{0.10\linewidth}rrrrrrrr}
    \hline
    Name  &  Visualization  &  $\mathrm{N_A}$  &  $\mathrm{N_B}$  &  L  &  C4  &  XACA  &  XACB  &  BpNP4CA  &  BpNP4CB  \\
    \hline
    Two-path &  \includegraphics[scale=\examplefigscale]{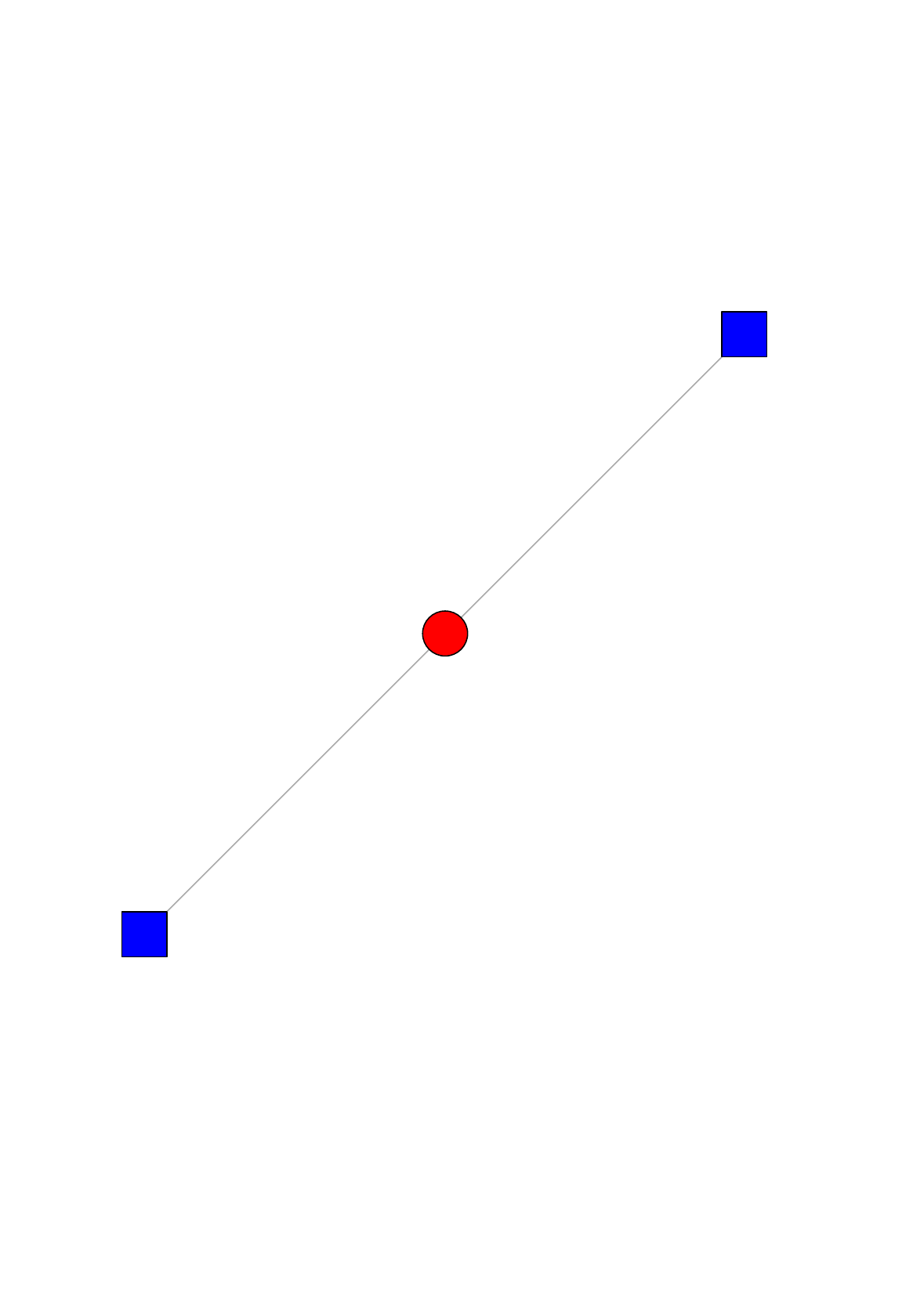}     & 1     & 2        & 2     & 0     & 1     & 0     & 0     & 0    \\
    Four-cycle & \includegraphics[scale=\examplefigscale]{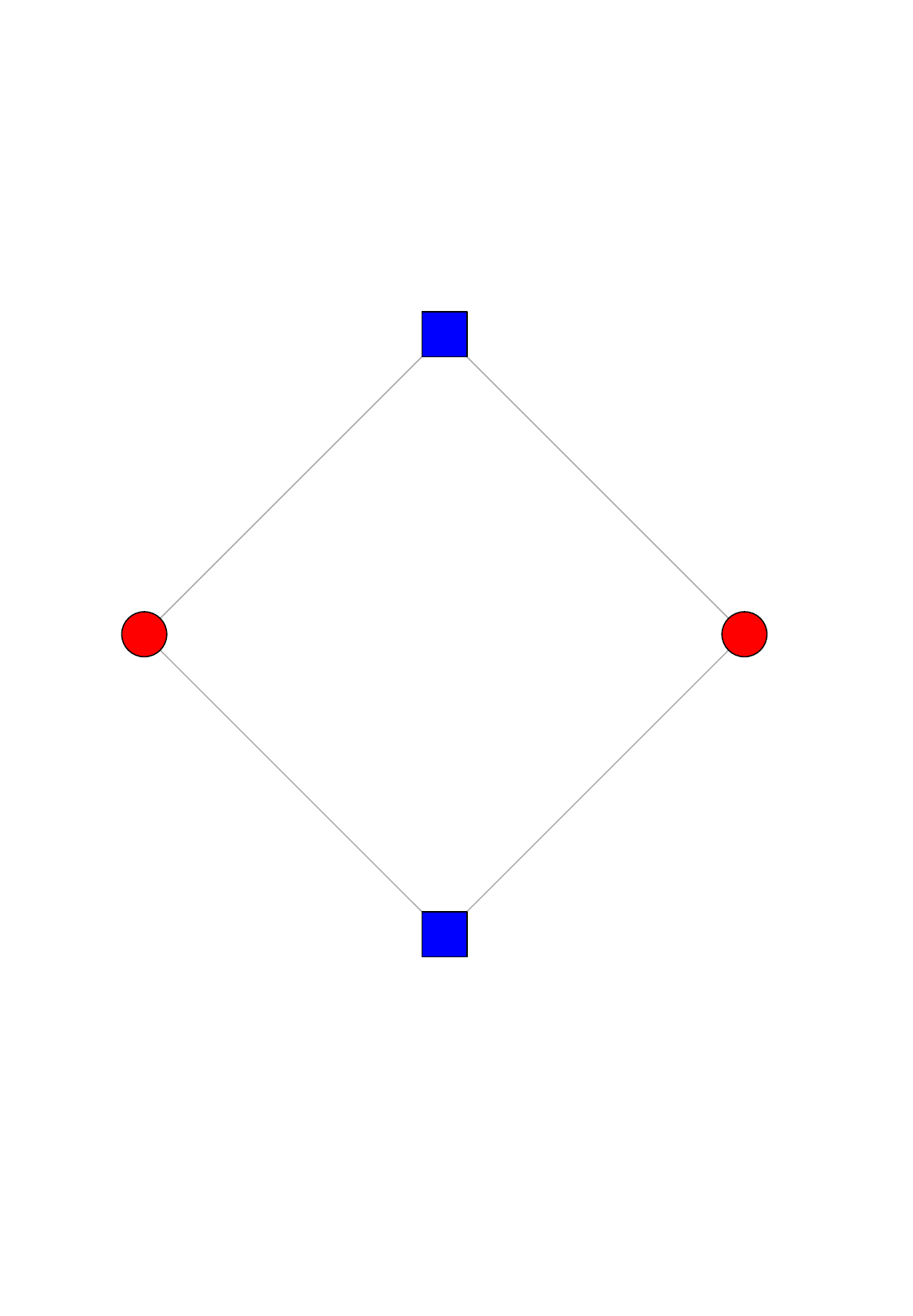}       & 2     & 2       & 4     & 1     & 1.5   & 1.5   & 2     & 2   \\
    Four-cycles-3 & \includegraphics[scale=\examplefigscale]{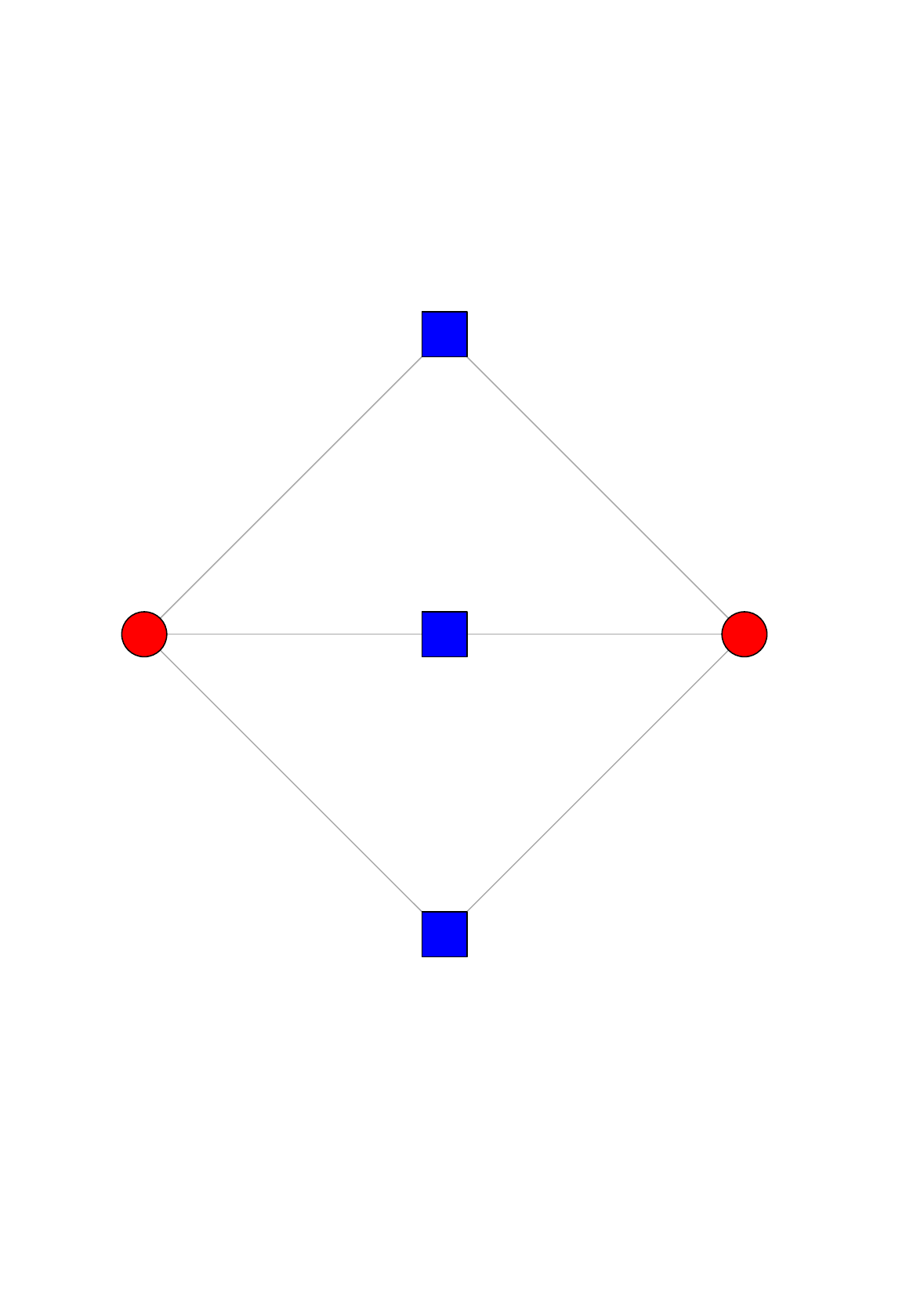}       & 2     & 3     & 6     & 3     & 4.5   & 1.75  & 3.4641 & 4.24264 \\
    Ten-cycle &    \includegraphics[scale=\examplefigscale]{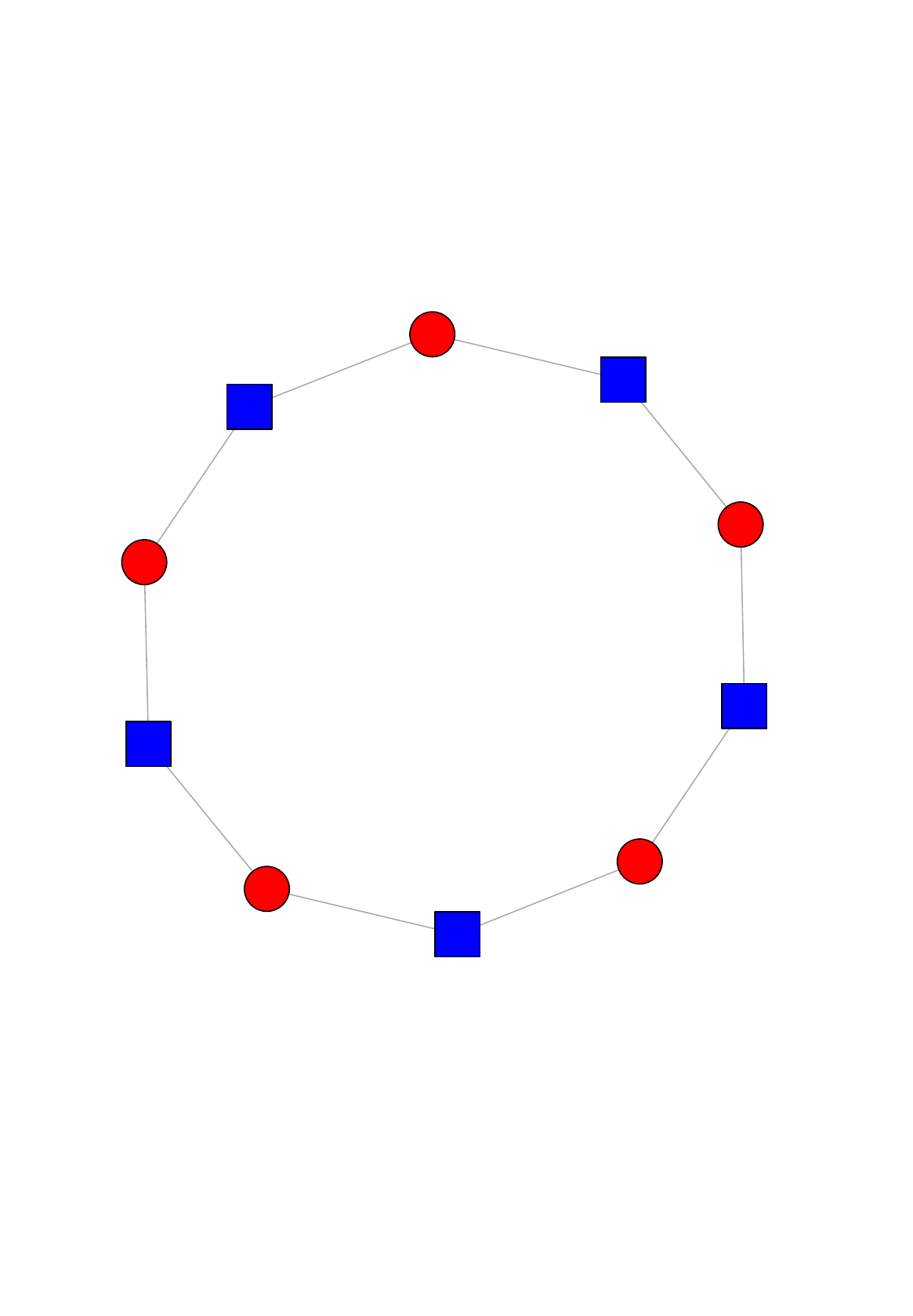}   & 5     & 5       & 10    & 0     & 5     & 5     & 0     & 0     \\
    Nine-star & \includegraphics[scale=\examplefigscale]{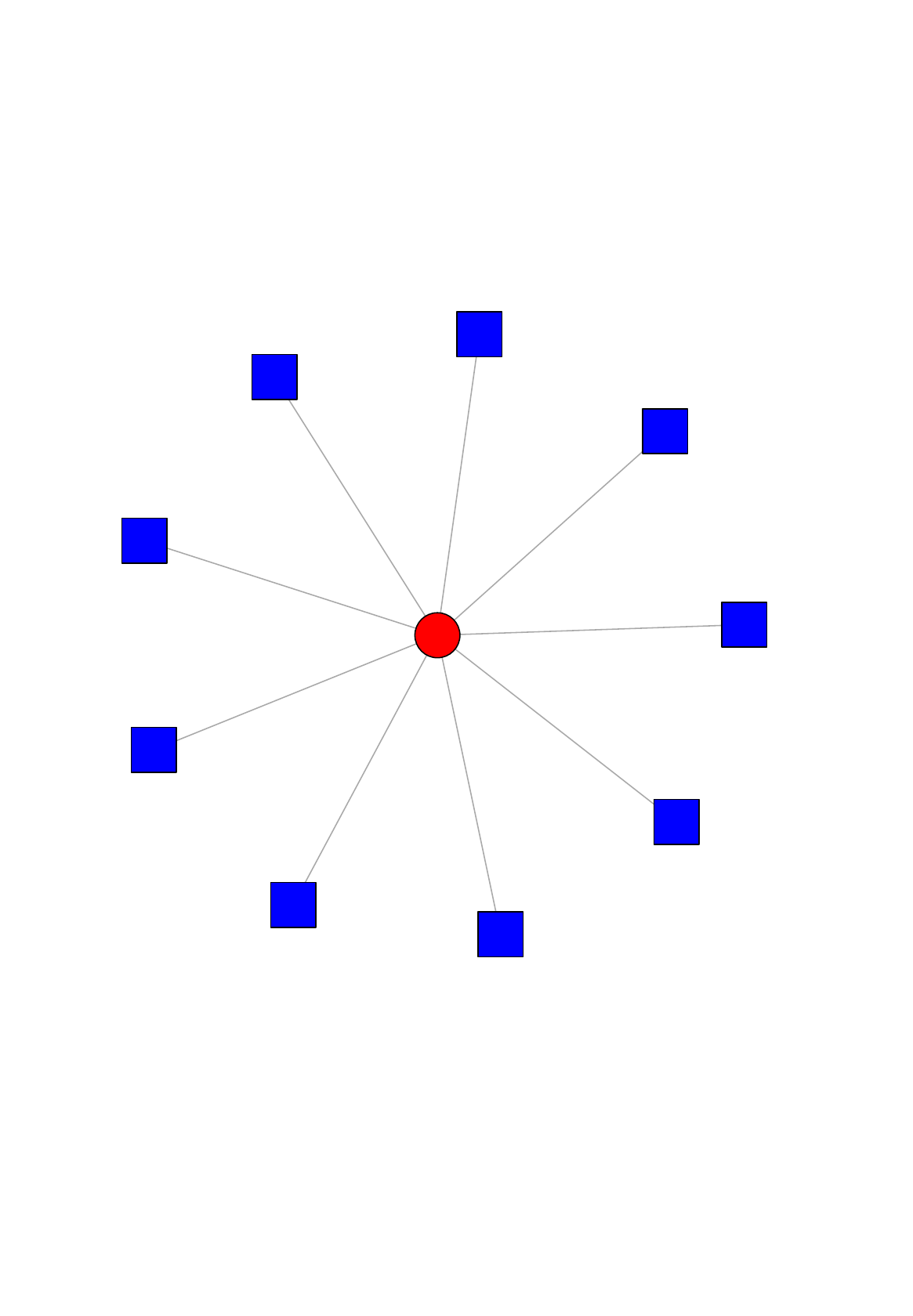}      & 1     & 9       & 9     & 0     & 36    & 0     & 0     & 0     \\
    Four-fan-3 & \includegraphics[scale=\examplefigscale]{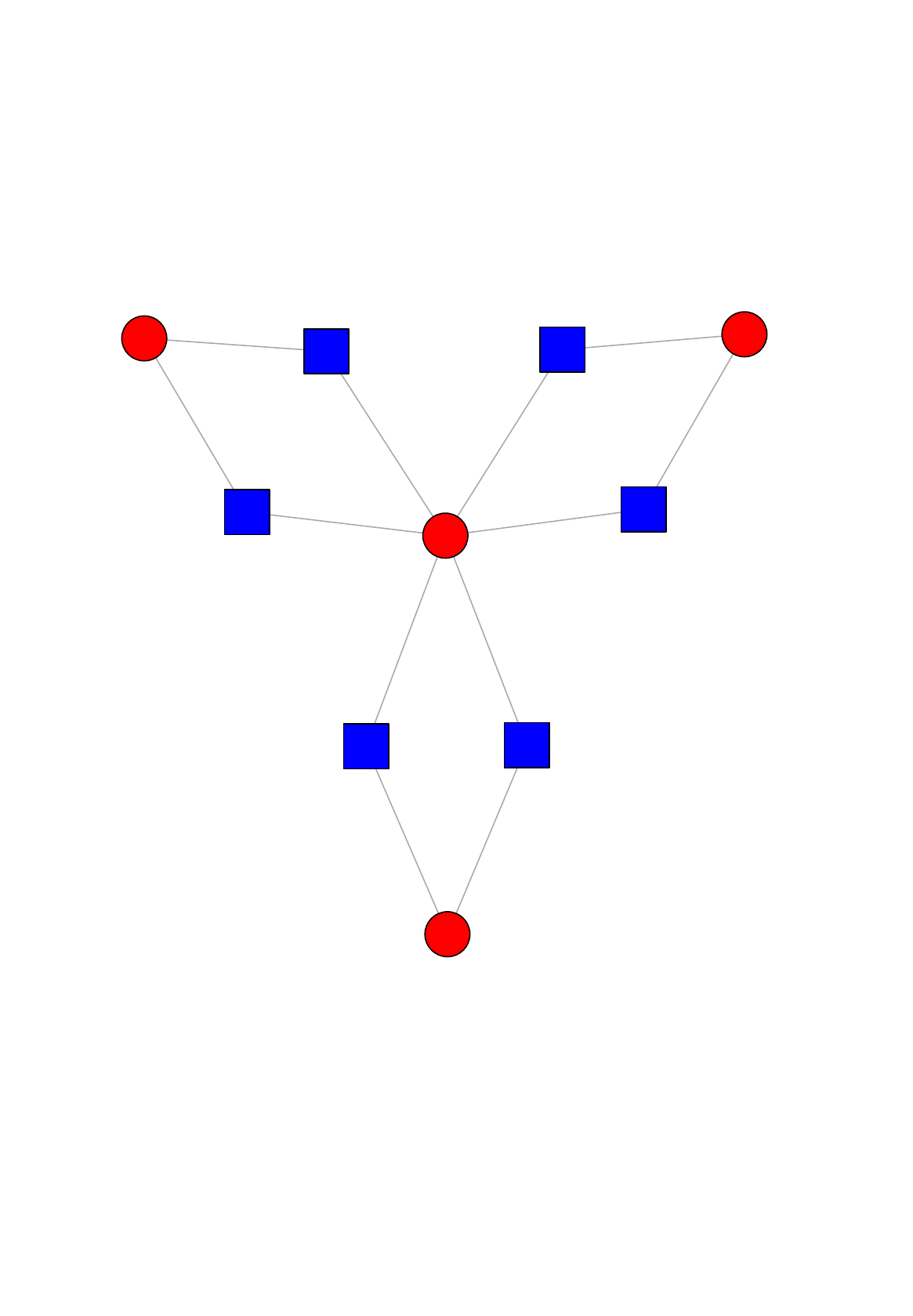}      & 4     & 6    & 12    & 3     & 16.5  & 4.5   & 4.73205 & 6 \\
    \hline
    \end{tabular}%
  \label{tab:examples}%
\end{table}%

In addition, Robins \etal{} \cite{robins07} report that interpretation of the alternating $k$-two-paths parameter is problematic:
``In this article, we do not concentrate on alternating $k$-two-path parameters. For some data,
we have found it important to include them in models but further work is needed to understand
better their effect when included with other parameters'' \cite[p.~201]{robins07}.
Martin \cite{martin20} described this as the authors ``being somewhat mystified by this statistic'' \cite[p.~86]{martin20}, and that most ERG modellers would not be able to describe a ``clear behavioral-process analogue to the once-canonical alternating two paths statistic'' \cite[p.~86]{martin20}.

Therefore, we propose new statistics that count four-cycles (and not
two-paths). In the following two sections we propose two different new
statistics. In Section~\ref{sec:altk4cycles} we propose a simple new
statistic based on \kca{} and \kcp{}. Unfortunately, however, this
statistic is shown to be more, rather than less, prone to degeneracy
than \kca{} and \kcp{}. Therefore, in Section~\ref{sec:newstats} we
propose a new statistic based on counting four-cycles, but which is
less prone to problems with near-degeneracy than the simple
four-cycles parameter or the \kca{} and \kcp{} parameters.

\section{A simple, but unsuccessful, new statistic}
\label{sec:altk4cycles}

A simple solution to the problem, described in
Section~\ref{sec:problems}, that the \kca{}
(\ref{qqn:bipartitealtkcyclesa}) and \kcp{} statistics count simple
two-paths as their first term ($k$-two-paths, $k=1$), is to remove the
first term and reverse the signs. In this way, the first, positive,
term no longer counts open two-paths, but rather counts four-cycles.
We define a
new statistic BipartiteAltK4CyclesA as
\begin{equation}
  \label{eqn:bipartitealtk4cyclesa}
   z_{\text{BipartiteAltK4CyclesA}(\lambda)} = -\left(  z_{\text{BipartiteAltKCyclesA}(\lambda)} - z_{\text{TwoPathsA}} \right),
\end{equation}
and its change statistic
\begin{equation}
  \label{ceqn:change_bipartitealtk4cyclesa}
  \delta_{\text{BipartiteAltK4CyclesA}(\lambda)}(i,j) = -\left( \delta_{\text{BipartiteAltKCyclesA}(\lambda)} - \graphdeg(i)   \right)
\end{equation}
where
\begin{equation}
  \label{eqn:twopaths}
  z_{\text{TwoPathsA}} = \sum_{i \in B} \sum_{\left\{l \in B\,:\,l<i\right\}} L_2(i, l)
\end{equation}
is the number of two-paths connecting nodes in node set $B$ (and which
therefore go through a node in node set $A$),
$L_2(i,l)$, defined by (\ref{eqn:L2}), is the number of two-paths connecting $i$ to $l$,
and $\graphdeg(i)$, the
degree of node $i$, is the change statistic for the number of
two-paths through node $i$.  BipartiteAltK4CyclesB and its change
statistic are defined similarly.

Table~\ref{tab:k4c_examples} is a copy of Table~\ref{tab:examples}, but with the new BipartieAltK4CyclesA and
BipartiteAltK4CyclesB statistics (labelled BpAK4CA and BpAK4CB
respectively) included (and the $\mathrm{N_A}$, $\mathrm{N_B}$, L,
BpNP4CA, and BpNP4CB columns removed to make space). This table
shows the value of the new statistics on some small example networks,
demonstrating, that, by design, they are zero for networks in which
there are no four-cycles.

\def \examplefigscale {0.09} 
\begin{table}[htbp!]
  \centering
  \caption{Statistics of some example bipartite networks. C4 is the number of four-cycles, and BpAK4CA and BpAK4CB are the new statistics BipartiteAltK4CyclesA and BipartiteAltK4CyclesB, respectively. Nodes in node set A are represented as red circles, and nodes in node set B as blue squares.}
    \begin{tabular}{lm{0.10\linewidth}rrrrr}
    \hline
    Name       &  Visualization                                                        & C4    &  XACA &  XACB &  BpAK4CA & BpAK4CB   \\
    \hline
    Two-path &  \includegraphics[scale=\examplefigscale]{twopath_bipartite.pdf}        & 0     & 1     & 0     &  0        &  0          \\
    Four-cycle & \includegraphics[scale=\examplefigscale]{fourcycle_bipartite.pdf}     & 1     & 1.5   & 1.5   &  0.5      &  0.5        \\
    Four-cycles-3 & \includegraphics[scale=\examplefigscale]{fourcycle3_bipartite.pdf} & 3     & 4.5   & 1.75  &  1.5      &  1.25       \\
    Ten-cycle &    \includegraphics[scale=\examplefigscale]{ring_bipartite.pdf}        & 0     & 5     & 5     &  0        &  0          \\
    Nine-star & \includegraphics[scale=\examplefigscale]{star_bipartite.pdf}           & 0     & 36    & 0     &  0        &  0          \\
    Four-fan-3 & \includegraphics[scale=\examplefigscale]{fourfan3_bipartite.pdf}      & 3     & 16.5  & 4.5   &  1.5      &  1.5        \\
    \hline
    \end{tabular}%
  \label{tab:k4c_examples}%
\end{table}%

\begin{figure}[ht!]
  \centering
  \includegraphics[width=\textwidth]{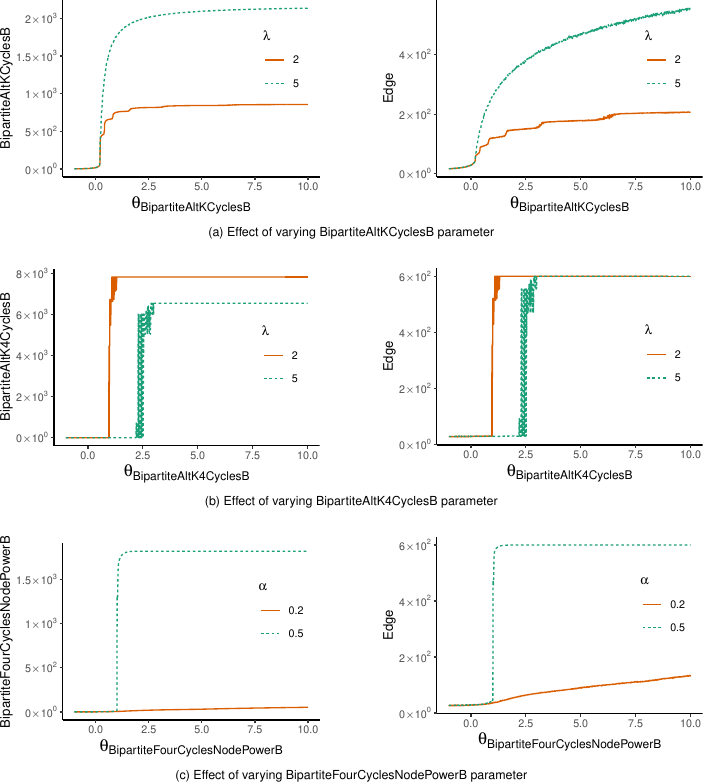}
  \caption{Effect of varying (a) the BipartiteAltKCyclesB parameter,
    (b) the BipartiteAltK4CyclesB parameter, and
    (c) the BipartiteFourCyclesNodePowerB parameter on the
    statistic corresponding to the parameter itself
    (left) and the Edge statistic (right). Each graph plots
    the mean value of the statistic (over 100 simulations) for two
    different values of the relevant $\lambda$ or $\alpha$ parameter.}
  \label{fig:altk4cycles_simulations}
\end{figure}

Unfortunately, however, simulation experiments indicate that the new
BipartiteAltK4CyclesA and BipartiteAltK4CyclesB parameters are actually
\emph{more} problematic with respect to near-degeneracy than the
original \kca{} and \kcp{}
parameters. Fig.~\ref{fig:altk4cycles_simulations} shows the results
of simulation experiments similar to those described by
Wang~\etal~\cite[p.~19]{wang09}. Bipartite networks were simulated with 30
nodes in node set $A$ and 20 nodes in node set $B$ with the Edge
parameter set to $-3.0$. In three different sets of simulations, for
each of the BipartiteAltKCyclesB, BipartiteAltK4CyclesB, and
BipartiteFourCyclesNodePowerB (defined in Section~\ref{sec:newstats}) parameters, the parameter in question is
varied from $-1.00$ to $10.0$ in increments of $0.01$ for each of two
values of $\lambda$ ($\lambda = 2$ and $\lambda = 5$) for
BipartiteAltKCyclesB and BipartiteAltK4CyclesB, and for each of two
values of $\alpha$ ($\alpha = 1/2$ and $\alpha = 1/5$) for
BipartiteFourCyclesNodePowerB.
The networks were simulated using the SimulateERGM program from the EstimNetDirected software package, using the basic ERGM sampler, with a burn-in of $10^5$ iterations and an interval of $10^4$ iterations between each of 100 samples, to ensure that samples are drawn from the equilibrium ERGM distribution, and are not too autocorrelated.

The right panel in Fig.~\ref{fig:altk4cycles_simulations}(a) shows
the results of a simulation similar to that shown in Fig,~9 of Wang
\etal~\cite{wang09}: the number of edges increases smoothly, giving
good coverage of the graph space (with respect to density, at
least). The left panel shows the value of the statistic
corresponding to the parameter BipartiteAltKCyclesB itself. The
behaviour of this curve is starting to look as if it could be prone to
near-degeneracy, with a fairly steep increase at a critical value.

The graphs in Fig.~\ref{fig:altk4cycles_simulations}(b) show the results for the new BipartiteAltK4CyclesB parameter. In this case, both the
Edge statistic (right) and BipartiteAltK4CyclesB statistic itself (left) show
a phase transition, where a critical value of the parameter
separates an empty graph regime from a complete graph regime. This model
is therefore near-degenerate, suggesting that this new parameter
may in fact be less useful than the original \kca{} and \kcp{} parameters.

For completeness, Fig.~\ref{fig:altk4cycles_simulations}(c)
shows the results for the new BipartiteFourCyclesNodePowerB
parameter defined in the following section (Section~\ref{sec:newstats}).
When $\alpha = 0.5$ an abrupt change from a near-empty to a full
graph occurs, however, as shown by the result for $\alpha = 0.2$, this
can be removed by decreasing the value of $\alpha$.

In summary, we defined new BipartiteAltK4CyclesA and BipartiteAltK4CyclesB parameters as modified forms of the \kca{} and \kcp{} parameters defined
in Wang~\etal~\cite{wang09}, in order to count four-cycles but
not open two-paths. However, near-degeneracy was exhibited in the simulation experiments (similar results occur with the larger simulated networks
described in Section~\ref{sec:simulations}; data not shown). We conclude, therefore, that although these new parameters
could potentially be useful in some cases, they are more prone to
near-degeneracy and not as useful as the existing \kca{} and
\kcp{} parameters. Therefore, in the following section, we define
new statistics that weight four-cycle counts in a different way.

\section{New statistics for modelling four-cycles in bipartite ERGMs}
\label{sec:newstats}

\begin{figure}
  \centering
  \includegraphics[width=\textwidth]{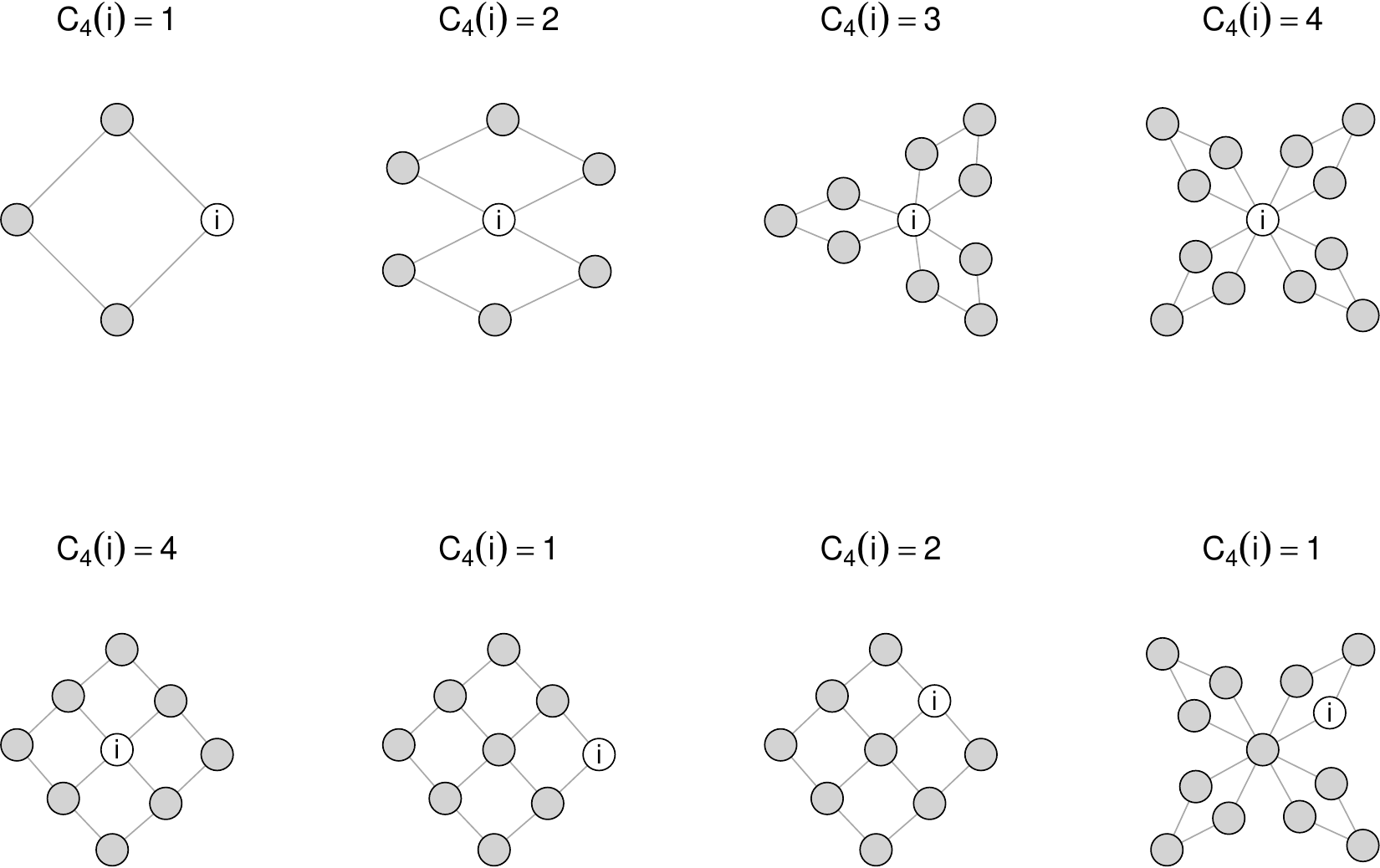}
  \caption{$C_4(i)$ is the number of four-cycles involving a node $i$.}
  \label{fig:c4i_examples}
\end{figure}

 Since
a four-cycle is a combination of two two-paths \cite[p.~123]{snijders06}, the
number of four-cycles is
\begin{equation}
  \label{eqn:C4}
  C_4 = \frac{1}{2}\sum_{i<j}\binom{L_2(i,j)}{2}.
\end{equation}
The sum in equation (\ref{eqn:C4}) is over the $\binom{n}{2}$ pairs of
nodes in the graph, with the factor of $\frac{1}{2}$ to account for
the double-counting due to the symmetry of each four-cycle containing
two distinct pairs of nodes, each connected by two two-paths.  The
number of four-cycles containing a particular node $i$ is
\begin{align}
  \label{eqn:nodeC4}
  C_4(i) &= \sum_{j \neq i}  \binom{L_2(i,j)}{2} \\
         &= \sum_{\left\{j\,:\, d(i,j) = 2 \right\}} \binom{L_2(i,j)}{2}
\end{align}
Some illustrative examples of the value of $C_4(i)$ for different nodes in some small graphs are shown in Fig.~\ref{fig:c4i_examples}.

The total
number of four-cycles (\ref{eqn:C4}) can also be expressed in terms of the
number of four-cycles at each node (\ref{eqn:nodeC4}) as
\begin{equation}
  C_4 = \frac{1}{4}\sum_i C_4(i)
\end{equation}
where the factor of $\frac{1}{4}$ accounts for the fact that each
four-cycle is counted four times, once for each node it contains.
The FourCyclesNodePower statistic is then defined as
\begin{equation}
  \label{eqn:FourCyclesNodePower}
  z_{\text{FourCyclesNodePower}(\alpha)} =  \sum_{i} \left[ C_4(i)^\alpha \right]
\end{equation}
where $0 < \alpha \leq 1$ is the exponent for, in the terminology
of Wilson \etal{}~\cite{wilson17}, the ``$\alpha$-inside'' weighting,
since the subgraph counts ($C_4(i)$ here) are exponentiated before
summing over all subgraphs. The ``$\alpha$-outside'' weighting would be
to exponentiate the statistic after summing over all subgraphs, that is,
in this case it would be $\left[ \sum_{i}  C_4(i) \right]^\alpha$.
As discussed in Wilson \etal{}~\cite{wilson17},
the $\alpha$-inside weighting leads to local dependence as usually
used in ERGMs, while the $\alpha$-outside weighting leads to global
dependence, in which all ties are dependent on each other to some degree
\cite[p.~41]{wilson17}. The nature of the local dependencies induced by the new change statistics defined here using the $\alpha$-inside weighting are discussed in Section~\ref{sec:dependence} below.

The statistics described in this section so far are equally applicable
to one-mode and two-mode (bipartite) graphs. When dealing with bipartite
graphs, however, it is often useful to consider statistics of the two
node sets separately. Hence we also define
\begin{equation}
  \label{eqn:BipartiteFourCyclesNodePowerA}
  z_{\text{BipartiteFourCyclesNodePowerA}(\alpha)} =  \sum_{i \in A} \left[ C_4(i)^\alpha \right]
\end{equation}
and
\begin{equation}
  \label{eqn:BipartiteFourCyclesNodePowerB}
  z_{\text{BipartiteFourCyclesNodePowerB}(\alpha)} =  \sum_{j \in B} \left[ C_4(j)^\alpha \right]
\end{equation}
for the two node sets $A$ and $B$ respectively. Because the sets $A$ and $B$ are disjoint, we have
\begin{equation}
  \label{eqn:nodepower_sum}
  z_{\text{FourCyclesNodePower}(\alpha)} = z_{\text{BipartiteFourCyclesNodePowerA}(\alpha)} + z_{\text{BipartiteFourCyclesNodePowerB}(\alpha)}.
\end{equation}

\begin{figure}
  \centering
  \includegraphics[width=\textwidth]{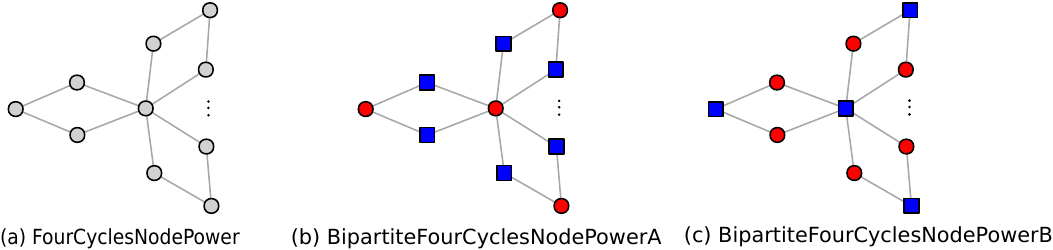}
  \caption{Representations of the new configurations
    (a) FourCyclesNodePower, (b) BipartiteFourCyclesNodePowerA, and
    (c) BipartiteFourCyclesNodePowerB. Nodes in node set A are shown as
    red circles, and nodes in node set B are shown as blue squares.}
  \label{fig:configurations}
\end{figure}

Representations of the new configurations are shown in
Fig.~\ref{fig:configurations}. The vertical ellipsis $\vdots$ in the
figures is to indicate that the configuration includes any number
(up to $\lfloor (n - 1) / 3 \rfloor$, since, apart from the one shared node, each four-cycle
must include at most three distinct nodes)
of
four-cycles all involving a shared node (the central node in the
figures).

Table~\ref{tab:examples} shows the values of the BipartiteFourCyclesNodePowerA (BpNP4CA) and BipartiteFourCyclesNodePowerB (BpNP4CB) statistics for some small example bipartite networks, with the parameter $\alpha = 0.5$ (the alternating $k$-two-path statistics XACA and XACB are shown with their parameter $\lambda = 2.0$). Note that, unlike the XACA and XACB (\kca{} and \kcp{}) statistics, these new statistics have the value zero for structures that contain no four-cycles (such as the two-path, ten-cycle, and nine-star structures in Table~\ref{tab:examples}).

The ``four-fan-3'' graph in Table~\ref{tab:examples} is the same graph
as the representation of the BipartiteFourCyclesNodePowerA
configuration in Fig.~\ref{fig:configurations}(b) (ignoring the vertical
ellipsis so that exactly three four-cycles are present). Note,
however, that this has the counterintuitive property that, although it
is a representation of the BipartiteFourCyclesNodePowerA configuration, in
fact the value of the BipartiteFourCyclesNodePowerB statistic is greater
than that of the BipartiteFourCyclesNodePowerA statistic for this graph. This is
because the value of the statistic
[(\ref{eqn:BipartiteFourCyclesNodePowerA}) or (\ref{eqn:BipartiteFourCyclesNodePowerB})] is the sum over all nodes in
the relevant node set ($A$ or $B$, respectively) of
the four-cycle count at each node (\ref{eqn:nodeC4}) raised to the power
$\alpha$ (``$\alpha$-inside'' weighting). Therefore, in this
graph, the nodes in mode $B$ contribute more to the total as
each one (of the six) is involved in exactly one four-cycle
(and hence raising to the power of $\alpha$ still contributes one to the sum),
while of the four nodes in mode $A$, three are involved in only one
four-cycle, while the fourth is involved in three four-cycles and
hence contributes only $3^\alpha \approx 1.73205$ (when $\alpha=0.5$).
Generalising this four-fan-3 graph to four-fan-$k$ ($k \geq 1$, and if $k=1$ the graph is just a four-cycle) with the central high-degree node in node set $A$, we have:
\begin{align}
  \label{eqn:four-fan=-j}
  N_A &= k+1 \\
  N_B &= 2k  \\
  n &= 3k+1 \\
  L &= 4k \\
  C_4 &= k \\
  \label{eqn:fourfank_bp4cnpA}
  z_{\text{BipartiteFourCyclesNodePowerA}(\alpha)} &= k^\alpha + k \\
  \label{eqn:fourfank_bp4cnpB}
  z_{\text{BipartiteFourCyclesNodePowerB}(\alpha)} &= 2k
\end{align}
Because $0 < \alpha \leq 1$, the BipartiteFourCyclesNodePowerA
statistic (\ref{eqn:fourfank_bp4cnpA}) will always be less than (or equal to, if  $\alpha = 1$ or $k = 1$) the
BipartiteFourCyclesNodePowerB statistic (\ref{eqn:fourfank_bp4cnpB})
for this family of graphs.

\subsection{Interpretation of the new parameters}
\label{sec:interpretation}

Interpretation of the FourCyclesNodePower parameter is that a positive value increases the number of four-cycles and a negative value decreases the number of four-cycles, relative to a value of zero. A smaller value of the exponent $\alpha$ means that additional four-cycles including the same node contribute less than if those cycles involved distinct nodes.

\begin{figure}
  \centering
  \includegraphics[width=\textwidth]{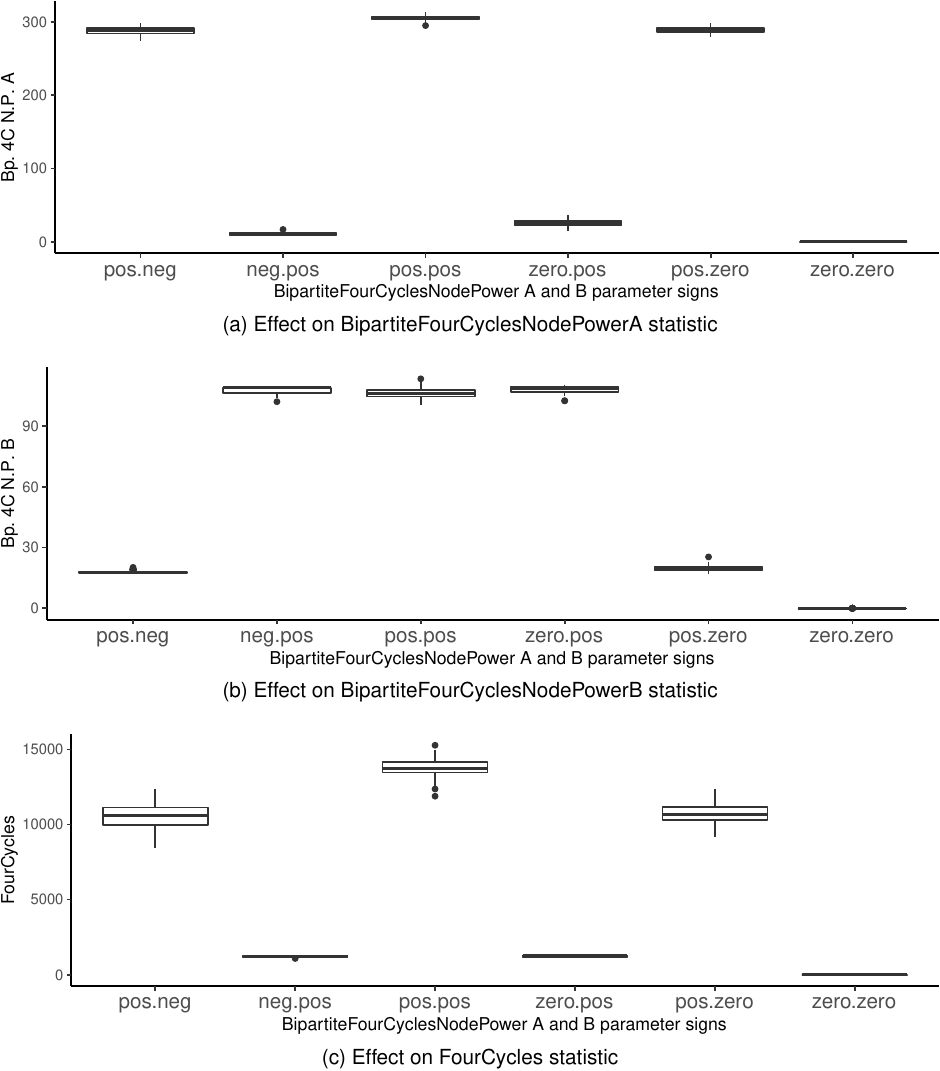}
  \caption{Effect of different combinations of negative, zero, and
    positive values of the BipartiteFourCyclesNodePowerA and
    BipartiteFourCyclesNodePowerB parameters on (a) the
    BipartiteFourCyclesNodePowerA statistic, (b) the
    BipartiteFourCyclesNodePowerB statistic and (c) the FourCycles
    statistic. Box plots show the statistics of 100 simulated
    networks.}
  \label{fig:simulation_boxplots}
\end{figure}

\begin{figure}
  \centering
  \includegraphics[width=\textwidth]{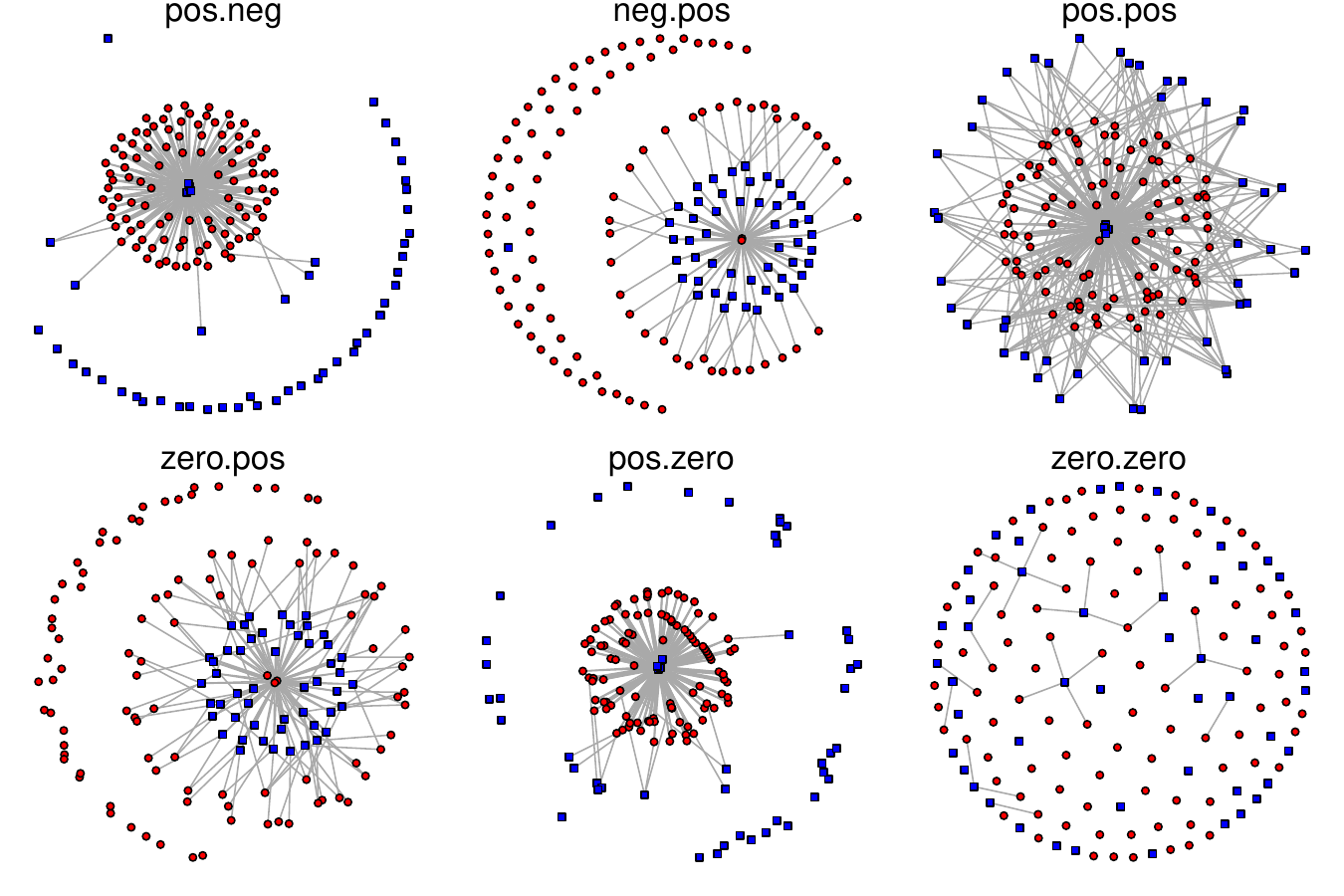}
  \caption{Examples of networks simulated with
    different combinations of negative, zero, and positive values of
    the BipartiteFourCyclesNodePowerA and
    BipartiteFourCyclesNodePowerB parameters, drawn from the simulations shown in
    Fig.~\ref{fig:simulation_boxplots}. Nodes in node set A are shown
    as red circles, and nodes in node set B are shown as blue
    squares.}
  \label{fig:simulation_visualizations}
\end{figure}

Interpretation of the BipartiteFourCyclesNodePowerA and
BipartiteFourCyclesNodePowerB parameters is rather more complicated
and is illustrated in Fig.~\ref{fig:simulation_boxplots} and
Fig.~\ref{fig:simulation_visualizations}.  These figures show
statistics (Fig.~\ref{fig:simulation_boxplots}) and network
visualizations (Fig.~\ref{fig:simulation_visualizations}) of simulated
bipartite networks with different combinations of positive, zero and
negative BipartiteFourCyclesNodePowerA and
BipartiteFourCyclesNodePowerB parameters.  The simulated networks have
100 nodes in node set $A$ and 50 nodes in node set $B$ and are
simulated with common ERGM parameters Edge, BipartiteAltStarsA
[$\lambda=2$], and BipartiteAltStarsB [$\lambda=2$] set to $-6.0$, $-0.4$,
and $1.0$, respectively.
For BipartiteFourCyclesNodePowerA [$\alpha=1/5$] and
BipartiteFourCyclesNodePowerB [$\alpha=1/5$] the negative
(``neg'') parameter value is $-1.5$ and the positive (``pos'')
parameter value is $6.5$.

Note that in a bipartite network, any four-cycle must contain two nodes in node set $A$ and two nodes in node set $B$. So how can we get more four-cycles in one mode than the other? The answer is that the four-cycle counts for the two must be equal, but the weighted node-oriented four-cycle counts (\ref{eqn:BipartiteFourCyclesNodePowerA}) and (\ref{eqn:BipartiteFourCyclesNodePowerB}) can differ. As discussed in Section~\ref{sec:newstats}, the statistic BipartiteFourCyclesNodePowerA (\ref{eqn:BipartiteFourCyclesNodePowerA}) is maximised by having four-cycles involving distinct pairs of nodes in node set $A$ (rather than many four-cycles involving the same node in node set $A$). If the BipartiteFourCyclesNodePower parameter for $A$ is positive and for $B$ is zero (or negative) then we tend to get more mode $A$ nodes involved in four-cycles, with the same mode $B$ nodes participating in many four-cycles, since the statistic is higher by having different nodes in the four-cycles, than for having the same node involved in many four-cycles. In the examples illustrated in Fig.~\ref{fig:simulation_visualizations}, this results in the mode $A$ nodes being part of a denser core with lots of four-cycles with a smaller number of $B$ nodes, resulting in isolated $B$ nodes. And vice versa for $A$ zero (or negative) and $B$ positive (``zero.pos'' and ``neg.pos''; these are perhaps clearer as there are more $A$ nodes than $B$ nodes in the network). Particularly in the ``neg.pos'' case, we can see a core of mode $B$ nodes connected to a small number of central mode $A$ nodes in four-cycles (as well as others not involved in four-cycles) and many mode $A$ isolates. If BipartiteFourCyclesNodePower for both $A$ and $B$ are positive then there are even more four-cycles, but they are more evenly distributed between the $A$ and $B$ nodes.

\begin{figure}[ht!]
  \centering
  \includegraphics[width=\textwidth]{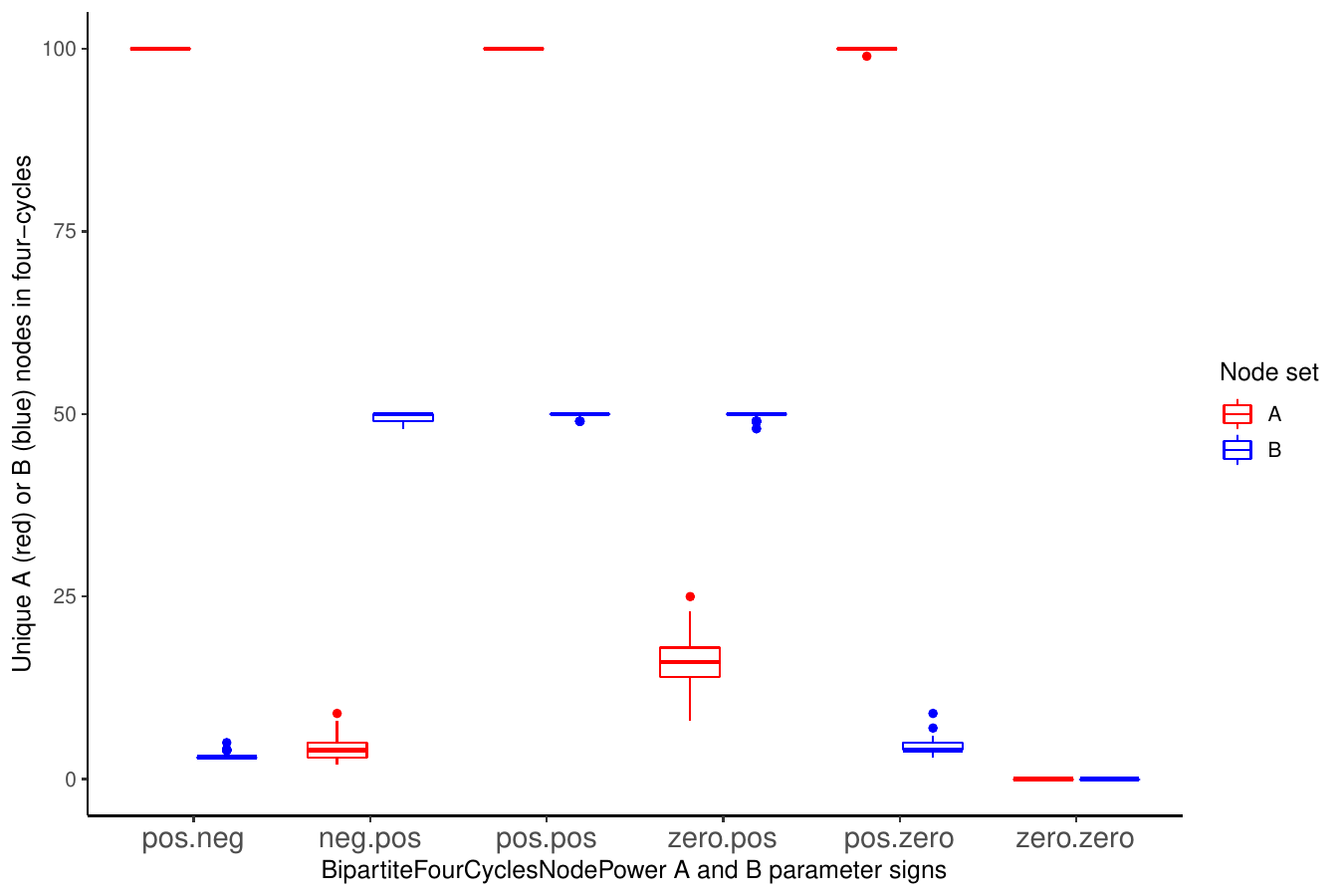}
  \caption{For each combination of negative, zero, and positive values
    of the BipartiteFourCyclesNodePowerA and
    BipartiteFourCyclesNodePowerB parameters, there are two box plots.
    They show the number of unique nodes in node set A (left, red) and
    node set B (right, blue) involved in four-cycles.  The simulations
    are the same as those used in Fig.~\ref{fig:simulation_boxplots},
    and as in that figure, combinations of only negative and zero
    values are not shown as they result in extremely low density graphs
    with no four-cycles (similar to the zero.zero case).}
  \label{fig:simulation_uniquenodes_boxplots}
\end{figure}

To try to make this interpretation clearer, consider
Fig.~\ref{fig:simulation_uniquenodes_boxplots}. The box plots in this
figure show the number of unique nodes in each node set ($A$ or $B$)
that are involved in four-cycles. When the
BipartiteFourCyclesNodePowerA parameter is positive and the
BipartiteFourCyclesNodePowerB parameter is zero or negative, then a
large number of node set $A$ nodes are involved in four-cycles, but
only a small number of node set $B$ nodes are. So the same node set
$B$ nodes are involved in multiple four-cycles with many different
node set $A$ nodes.  If BipartiteFourCyclesNodePowerB is positive and
BipartiteFourCyclesNodePowerA is negative or zero, then the same nodes
in mode set $A$ are involved in multiple four-cycles with many
different nodes from node set $B$.

Equation (\ref{eqn:nodepower_sum}) implies that there are two degrees of freedom for the three parameters; for example we can include both BipartiteFourCyclesNodePowerA and BipartiteFourCyclesNodePowerB in a model, but not also FourCyclesNodePower since it is the sum of the other two.

\subsection{Change statistics for the new statistics}
\label{sec:changestats}

The change statistic \cite{hunter06,snijders06,hunter12}, that is, the difference in the statistic caused by adding a new edge $(i,j)$, for the four-cycles statistic (\ref{eqn:C4}) is
\begin{align}
  \label{eqn:delta_C4}
  \delta_{C4}(i,j) &= \sum_{k \in N(i)} L_2(j, k) \\
                   &= \sum_{k \in N(j)} L_2(i, k).
\end{align}
The change statistic for the FourCyclesNodePower
statistic (\ref{eqn:FourCyclesNodePower}) is then:
\begin{equation}
  \label{eqn:delta_FourCyclesNodePower}
  \begin{split}
    \delta_{\text{FourCyclesNodePower}(\alpha)}(i, j)  & = \left[ C_4(i) + \delta_{C4}(i,j) \right]^\alpha - C_4(i)^\alpha \\
    & + \left[ C_4(j) + \delta_{C4}(i,j) \right]^\alpha - C_4(j)^\alpha \\
    & + \sum_{k \in N(i)} \left[ \left( C_4(k) + L_2(k, j) + x_{kj}L_2(k,i) \right)^\alpha - C_4(k)^\alpha \right] \\
    & + \sum_{\left\{ k \,:\, k \in N(j) \land\, k \not\in N(i) \right\}} \left[ \left( C_4(k) + L_2(k, i) \right)^\alpha - C_4(k)^\alpha \right].
  \end{split}
\end{equation}
The four terms in equation (\ref{eqn:delta_FourCyclesNodePower}) count
the contributions from, respectively, node $i$, node $j$, the
neighbours of node $i$, and the neighbours of node $j$ which are not
also neighbours of node $i$.  Note that in the third term (the
contribution from neighbours of node $i$), a node $k$ can only be a
neighbour of both node $i$ and node $j$ (that is, $k \in N(i) \land
x_{kj}=1$) if the network is not bipartite.

The change statistic for the bipartite four-cycles statistic for the
node set $A$ (\ref{eqn:BipartiteFourCyclesNodePowerA}) is simpler than
the general case (\ref{eqn:delta_FourCyclesNodePower}), as we only
count the contributions from the nodes in node set $A$. Specifically,
we have:
\begin{equation}
  \label{eqn:delta_BipartiteFourCyclesNodePowerA}
  \begin{split}
    \delta_{\text{BipartiteFourCyclesNodePowerA}(\alpha)}(i, j) & = \left[ C_4(i) + \delta_{C4}(i, j) \right]^\alpha - C_4(i)^\alpha \\
    & + \sum_{k \in N(j)} \left[ \left( C_4(k) + L_2(k, i) \right)^\alpha - C_4(k)^\alpha \right]
  \end{split}
\end{equation}
where $i \in A$, $j \in B$, and $k \in A$. The change statistic for
BipartiteFourCyclesNodePowerB
(\ref{eqn:BipartiteFourCyclesNodePowerB}) is defined analogously.

\subsection{Position of the new statistics in the dependence hierarchy}
\label{sec:dependence}

\begin{figure}
  \centering
  \includegraphics[width=.8\textwidth]{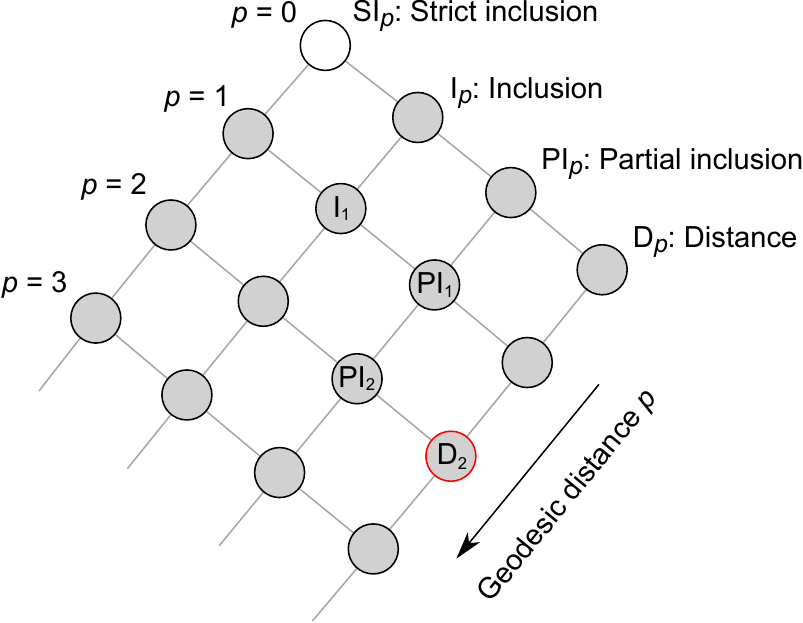}
  \caption{Dependence hierarchy, adapted from Wang \etal{}
    \cite[Fig.~3]{wang13_bipartite}. $\mathrm{SI}_0$ is not defined.}
  \label{fig:hierarchy}
\end{figure}

The configurations allowed in a model are determined by the assumptions as to which ties are allowed to depend on which other ties. Pattison \& Snijders \cite{pattison13_in_lusher13book} (subsequently elucidated by Wang \etal{}~\cite{wang13_bipartite} for bipartite networks, and more recently by Pattison \etal{}~\cite{pattison24}) created a two-dimensional hierarchy of dependence assumptions, where the two dimensions are two facets of proximity: the form of the proximity condition, and the maximum distance between dependent ties. This two-dimensional hierarchy of dependence assumptions is illustrated in Fig.~\ref{fig:hierarchy}, showing a partial order structure, in which, if one dependence condition can be implied by another, it can be reached by a downwards path from the first to the second \cite[p.~215]{wang13_bipartite}.

In order to describe the proximity conditions, it is useful to define some notation for neighbourhoods in a graph. In 
Section~\ref{sec:ergm} we defined $N(i)$ as the neighbours of node $i$, that is, nodes
$k \neq i$ such that $d(i,k) = 1$. Following Pattison \etal{}~\cite{pattison24}, we now define
$N_q(u)$, the $q$-neighbourhood of node $u$, as the set of nodes within geodesic distance $q$ of $u$, that is,
nodes $v$ such that $d(u, v) \leq q$. 
Note that $N(u)$ as earlier defined in Section~\ref{sec:ergm} is distinct from $N_1(u)$, the former being defined as nodes with a geodesic distance of exactly 1 from $u$ (and hence excluding $u$ itself), while the latter includes $u$ itself as $d(u,u)=0$.
Further, we define $N_q(U)$, the $q$-neighbourhood of node set $U$, as
$N_q(U) = \{v: v \in N_q(u) \text{ for some } u \in U\}$, that is, the set of nodes whose distance to some node in $U$ is no
more than $q$.

The four forms of proximity conditions, describing the nature of the proximity between the neighbourhoods of two pairs of nodes whose respective tie variables are hypothesized to be conditionally dependent only if the proximity condition holds, can be summarised as follows \cite{pattison24}, in order of increasing generality:
\begin{enumerate}
    \item Strict $p$-inclusion. $\mathrm{SI}_p$  ($p \geq 1$) holds if the $p$-neighbourhood of each node in a pair includes both of the nodes in the other pair.
    \item $p$-inclusion. $\mathrm{I}_p$ ($p \geq 0$) holds if the $p$-neighbourhood of each pair of nodes includes the other pair.
    \item Partial $p$-inclusion. $\mathrm{PI}_p$ ($p \geq 0$) holds if the $p$-neighbourhood of one pair of nodes includes the other pair.
    \item $p$-proximity. $\mathrm{D}_p$ ($p \geq 0$) holds if the $p$-neighbourhood of one pair of nodes has a non-empty intersection with the other pair.
\end{enumerate}

The dependence condition $\mathrm{I}_1$ is equivalent to the widely-used ``social circuit`` dependence assumption \cite{wang13_bipartite}, in which two ties are conditionally dependent if they form a four-cycle if both present 
\cite{pattison02,pattison04,snijders06,koskinen13}, and $\mathrm{PI}_1$ allows the ``alternating pendant-triangle'' statistics recently described by Pattison \etal{}~\cite{pattison24}.

The simple but unsuccessful new statistics described in
Section~\ref{sec:altk4cycles} are in the same dependence class
($\mathrm{I}_1$, ``social circuit'') as the original \kca{} and \kcp{}
statistics from which they are derived.

The change statistic for FourCyclesNodePower
(\ref{eqn:delta_FourCyclesNodePower}) in computing the probability of
a new edge $(i,j)$, depends only on edges between nodes in the
two-neighbourhoods of $i$ and of $j$, since any two nodes in a
four-cycle must be at a geodesic distance of at most two from each
other (a four-cycle is a pair of nodes with two two-paths between
them). Hence any edges on which the probability of edge $(i,j)$ depend
must be in the two-neighbourhood of $\{i,j\}$.
This puts the
FourCyclesNodePower configuration in the dependence class two-proximity
($\mathrm{D}_2$): the $p$-neighbourhood (with $p=2$) of
one pair of nodes has a non-empty intersection with the other pair
\cite{pattison13_in_lusher13book,pattison24}.
$\mathrm{D}_2$ is labelled with a red outline in the dependence
hierarchy diagram shown in Fig.~\ref{fig:hierarchy}.

Because of the partial order structure of the dependence hierarchy,
the stricter proximity forms for a fixed $p$ imply the more general
ones (and $\mathrm{D}_p$ is the most general), and for a fixed
proximity condition, smaller $p$ implies all the larger $p$ (so
$\mathrm{D}_1$ implies $\mathrm{D}_2$ for instance). Hence, to show
that the FourCyclesNodePower configuration, which is in $\mathrm{D}_2$, is
not also in any more specific dependence class, it is sufficient to show
that it is not in $\mathrm{D}_1$ and also not in $\mathrm{PI_2}$ (see
Fig.~\ref{fig:hierarchy}). To do so, we can use
Proposition~3 of Pattison \etal{}~\cite{pattison24}, which gives
the properties that must hold for configurations implied by
the dependence structures associated with each proximity condition.

Proposition 3(d) of Pattison \etal{}~\cite{pattison24} states that
``For $\mathrm{D}_\mathrm{p}$, each configuration is a subgraph in which
every pair of edges lies on a path of length $\leq (p+2)$''
\cite[p.~191]{pattison24}. So for a configuration to be in
$\mathrm{D}_1$, every pair of edges must lie on a path of length
three (or shorter).
The FourCyclesNodePower configuration
(Fig.~\ref{fig:configurations}(a)) does not meet this requirement, since
there are edges that do not lie on a path of length three or less:
consider, for example, a pair of edges incident to the outermost node
in the figure on two different four-cycles. These do not lie on a path
of length three (but are on a path of length four, satisfying the
requirement for $\mathrm{D}_2$ but not $\mathrm{D}_1$).
Hence the
FourCyclesNodePower configuration is not in dependence class $\mathrm{D}_1$.

Proposition 3(c) of Pattison \etal{}~\cite{pattison24} states that
``For $\mathrm{PI}_\mathrm{p}$, each configuration is a subgraph in which every pair
of edges lies either on a cyclic walk of length $\leq (2p+2)$ or on
a cyclic walk of length $\leq 2(p-r)+1$ with an additional path of length
$\leq r+1$ attached to a node lying on the cyclic walk, for
$0 \leq r \leq p-1$'' \cite[p.~191]{pattison24}.
So, for $\mathrm{PI}_2$, each configuration is a subgraph in
which every pair of edges is on a cyclic walk of length $\leq (2p+2) =
6$, or on a cyclic walk of length $\leq 2(p-0)+1 = 5$ with an
additional path of length $\leq 1$ attached to a node on the cyclic
walk, or on a cyclic walk of length $\leq 2(p-1)+1 = 3$ with an
additional path of length $\leq 2$ attached to a node on the cyclic
walk. Again, we can see that the configuration for FourCyclesNodePower
does not meet these conditions, considering a pair of maximally distant
edges (those incident to the outermost nodes in two different four-cycles in Fig.~\ref{fig:configurations}(a)). Such a pair of edges is neither on a six-cycle,
and nor is it on a five-cycle with an additional path of length  at most one
or a three-cycle with an additional path of length at most two. Hence the FourCyclesNodePower configuration is not in dependence class $\mathrm{PI}_2$.

In the case of the BipartiteFourCyclesNodePower ($A$ and $B$)
statistics for bipartite networks, the same reasoning applies (see
Fig.~\ref{fig:configurations}(b) and Fig.~\ref{fig:configurations}(c)).

\subsection{Implementation}
\label{sec:implementation}

The new ERGM effects FourCyclesNodePower, BipartiteFourCyclesNodePowerA, and BipartiteFourCyclesNodePowerB are implemented in the EstimNetDirected \cite{stivala20} software, available from \url{https://github.com/stivalaa/EstimNetDirected}.
BipartiteFourCyclesNodePowerA and BipartiteFourCyclesNodePowerB are also implemented, as b1np4c and b2np4c, as user-contributed statnet model terms
\cite{hunter13,hunter19}, available from \url{https://github.com/stivalaa/ergm.terms.contrib}. An example is described in Section~\ref{sec:statnet_example}.

To count the number of unique type $A$ and $B$ nodes that are involved in four-cycles in a two-mode network, the CYPATH
software (\url{http://research.nii.ac.jp/~uno/code/cypath.html})
\cite{uno14} was used to enumerate all of the four-cycles (which are necessarily chordless in a bipartite network).

Scripts for data conversion, statistical analysis, and generating plots were written in R \cite{R-manual} using the igraph \cite{csardi06,antonov23} and ggplot2 \cite{ggplot2} packages.

\section{Simulation experiments}
\label{sec:simulations}

In order to investigate the effect of the BipartiteFourCyclesNodePowerA parameter, and compare it to that of the \kca{} \cite{wang09} (known as XACA in MPNet and BipartiteAltKCyclesA in EstimNetDirected) parameter, we conducted some simulation experiments. In these experiments, bipartite networks with 750 nodes in node set $A$ and 250 nodes in set $B$ were simulated with the Edge, BipartiteAltStarsA [$\lambda=2$], and BipartiteAltStarsB [$\lambda=2$] parameters set to $-8.50$, $-0.20$, and $2.00$, respectively. In one set of experiments, the BipartiteAltKCyclesA parameter was varied from $-1.00$ to $1.00$ in increments of $0.01$, for each of three values of $\lambda$: $2$, $5$, and $10$. In another set of experiments, the BipartiteFourCyclesNodePowerA parameter was varied from $-1.00$ to $2.00$ in increments of $0.01$, for each of three values of $\alpha$: $1/10$, $1/5$, and $1/2$.
The networks were simulated using the SimulateERGM program from the EstimNetDirected software package, using the tie/no-tie (TNT) sampler \cite{morris08}, with a burn-in of $10^7$ iterations and an interval of $10^5$ iterations between each of 100 samples, to ensure that samples are drawn from the equilibrium ERGM distribution, and are not too autocorrelated.

\begin{figure}
  \centering
  \includegraphics[width=\textwidth]{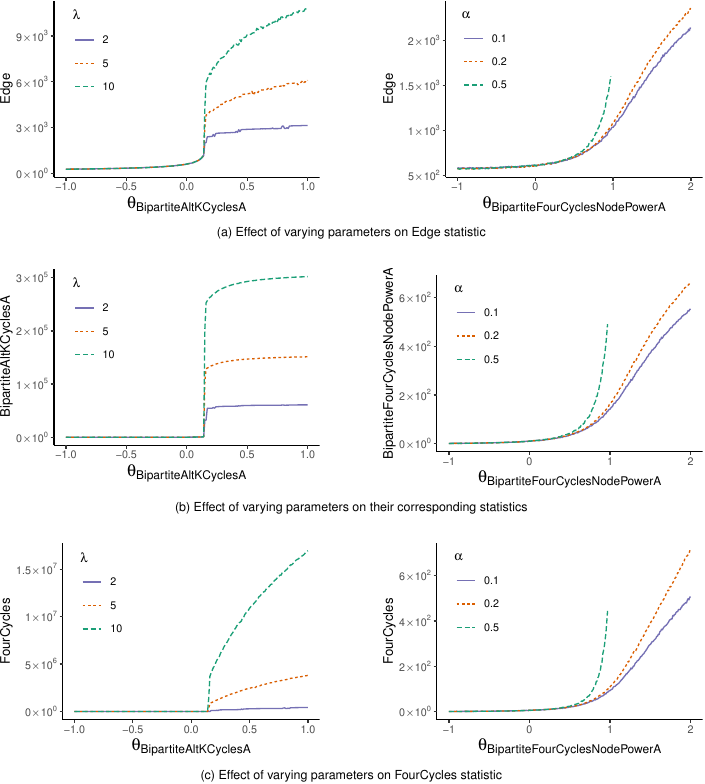}
  \caption{Effect of varying the BipartiteAltKCyclesA parameter (left)
    and BipartiteFourCyclesNodePowerA parameter (right) on (a) the Edge statistic,
    (b) the statistic corresponding to the parameter itself
    and (c) the FourCycles statistic. Each graph plots
    the mean value of the statistic (over 100 simulations) for three
    different values of the $\lambda$ or $\alpha$ parameter for
    BipartiteAltKCyclesA and BipartiteFourCyclesNodePowerA,
    respectively}
  \label{fig:simulations}
\end{figure}

The results of these simulations are shown in Fig.~\ref{fig:simulations}. Using the BipartiteAltKCyclesA parameter (left column) results in phase transition or near-degeneracy behaviour, with the statistic showing a sudden sharp increase at a critical value of the parameter. At this critical value, the graph density (Fig.~\ref{fig:simulations}(a) left plot) also sharply increases, as does the number of four-cycles (Fig.~\ref{fig:simulations}(c) left plot), which, at parameter values less than the critical value, hardly increased at all. This behaviour is similar to that of the simple Markov (edge-triangle) model described in Koskinen \& Daraganova \cite{koskinen13}, and is characteristic of near-degeneracy in ERGMs. This can prevent estimation of models which contain parameters that cause this behaviour, and yet occurs in this case even when using an ``alternating'' statistic, designed to try to avoid such behaviour \cite{wang09}. Note that changing the $\lambda$ parameter appears merely to change the maximum value of the statistic; it does not remove or ``smooth out'' the phase transition. In contrast, when using the new BipartiteFourCyclesNodePowerA parameter (right column) the phase transition behaviour is less apparent even for the highest value of $\alpha$ ($1/2$), and can be smoothed out further as we decrease $\alpha$. So, by appropriately setting $\alpha$, we can use the BipartiteFourCyclesNodePowerA parameter to generate smoothly varying numbers of four-cycles, without suddenly tipping from a low-density low-clustering regime to a high-density high-clustering regime with nothing in between, which did not seem to be possible on this example with the BipartiteAltKCyclesA parameter.

\section{Empirical example}
\label{sec:statnet_example}

The new statistics BipartiteFourCyclesNodePowerA and
BipartiteFourCyclesNodePowerB were implemented in statnet as b1np4c and
b2np4c, using the facility to define custom ergm model terms
\cite{hunter13,hunter19}. They are available at
\url{https://github.com/stivalaa/ergm.terms.contrib}.

Here we demonstrate this implementation by using the the statnet ergm
package \cite{ergm} with the new user terms to estimate a model for
the Davis ``Southern Women'' network \cite{davis41}, obtained via the
latentnet R package \cite{krivitsky08,latentnet}. The network
represents the participation of 18 women (first mode) in 14 social
events (second mode).
This is a well-known affiliation network \cite{lerner22}, having been
used as an example by Breiger~\cite{breiger74} and many papers since,
including two \cite{wang09,kevork22} in the literature survey
(Table~\ref{tab:literature}). Although the original publication
\cite{davis41} contains information on the event times, with some
exceptions \cite{everett18,lerner22} this is not usually used, and we do not
use this information here.

The b1np4c and b2np4c model terms take the $\alpha$ ($0 < \alpha \leq 1$)
value as a parameter, with a default value of $\alpha = 0.5$ if omitted.
For example, to estimate a model with the b2np4c (BipartiteFourCyclesNodePowerB) term  with $\alpha = 1/5$ (Model~4 in Table~\ref{tab:southern_women_ergm_results}):
\begin{verbatim}
davis_model4 <- ergm(davis ~ edges + gwb1degree(1, TRUE) +
                             gwb2degree(1, TRUE) + b2np4c(1/5),
  control = control.ergm(main.method = "Stochastic-Approximation"))
\end{verbatim}

\begin{table}[!htbp]
  \caption{ERGM parameter estimates for the Southern Women network,
    estimated with statnet. The table was generated directly from the statnet models with the texreg R package \cite{leifeld13}.}
  \label{tab:southern_women_ergm_results}  

{\begin{tabular*}{\textwidth}{@{\extracolsep{\fill}}lcccc@{}}        
\hline
 & Model 1 & Model 2 & Model 3 & Model 4 \\
\hline
edges             & $-2.07^{***}$ & $-0.20$      & $0.47$        & $-5.90^{***}$ \\
                  & $(0.34)$      & $(0.24)$     & $(0.31)$      & $(0.58)$      \\
b1star2           & $0.07$        &              &               &               \\
                  & $(0.07)$      &              &               &               \\
b2star2           & $0.18^{***}$  &              &               &               \\
                  & $(0.04)$      &              &               &               \\
gwb1deg.fixed.1   &               & $-0.83$      & $-7.04^{***}$ & $10.60^{***}$ \\
                  &               & $(1.03)$     & $(1.51)$      & $(1.80)$      \\
gwb2deg.fixed.1   &               & $-2.26^{**}$ & $7.76^{**}$   & $-6.95^{***}$ \\
                  &               & $(0.85)$     & $(2.48)$      & $(1.57)$      \\
gwb1dsp.fixed.0.5 &               &              & $0.45^{***}$  &               \\
                  &               &              & $(0.12)$      &               \\
gwb2dsp.fixed.0.5 &               &              & $-1.33^{***}$ &               \\
                  &               &              & $(0.30)$      &               \\
b2np4c.fixed.0.2  &               &              &               & $17.30^{***}$ \\
                  &               &              &               & $(1.86)$      \\
\hline
AIC               & $319.34$      & $328.85$     & $308.77$      & $285.41$      \\
BIC               & $329.93$      & $339.44$     & $326.41$      & $299.53$      \\
Log Likelihood    & $-156.67$     & $-161.43$    & $-149.38$     & $-138.71$     \\
\hline
\multicolumn{5}{l}{\scriptsize{$^{***}p<0.001$; $^{**}p<0.01$; $^{*}p<0.05$}}
\end{tabular*}}{}
\end{table}

Table~\ref{tab:southern_women_ergm_results} shows four
models for the Southern Women network, estimated with the stochastic approximation algorithm \cite{snijders02}. Model~1, with only the edges and two-star terms for each mode,
is the same as \cite[Model~(8.3)]{wang09}. Model~2, using the
geometrically weighted degree terms rather than two-stars, is similar
to \cite[Model~(8.5)]{wang09}, but using the statnet gwb1degree and
gwb2degree terms rather than the BPNet alternating $k$-star terms
\ksp{} and \ksa{}. Note the reversal of interpretation of signs
between \ksp{}/\ksa{} and gwb1degree/gwb2degree
\cite{levy16poster,levy16}.  Model~3, adding the geometrically
weighted dyadwise shared partner terms, is similar to
\cite[Model~(8.6)]{wang09}.  Model~4 uses the new b2np4c term rather than
the geometrically weighted dyadwise shared partner terms gwb1dsp and
gwb2dsp (models with b1np4c did not converge).

\begin{figure}
  \centering
  \includegraphics{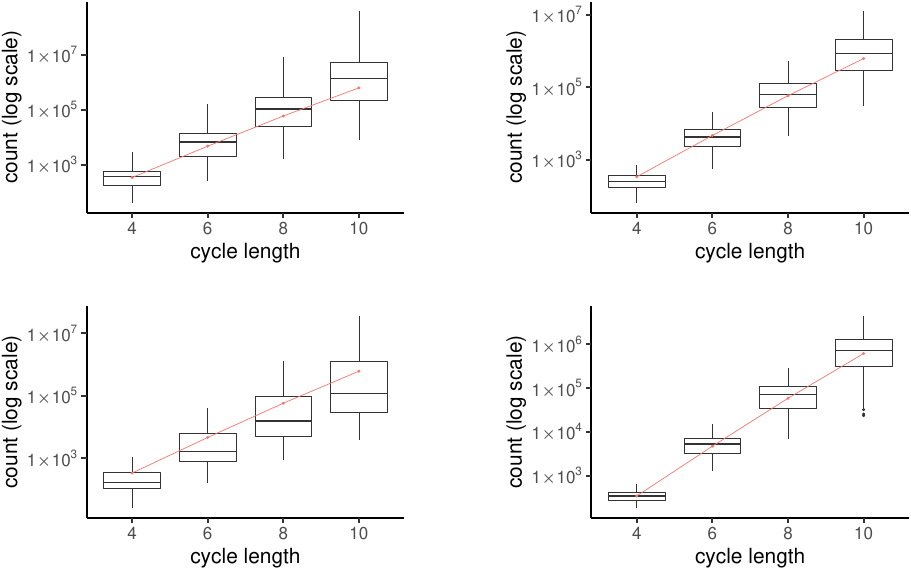}
  \caption{Cycle length distribution goodness-of-fit plots for the
    Southern Women network ERGM (Table~\ref{tab:southern_women_ergm_results})
    Model 1 (top left), Model 2 (top right),
    Model 3 (bottom left), and Model 4 (bottom right). Observed
    network statistics are plotted as red points (joined by red lines
    as a visual aid) with the statistics of 100 simulated networks
    plotted as black box plots.}
  \label{fig:southern_women_cycledist_gof_plots}
\end{figure}

Cycle length distribution goodness-of-fit plots for the four models
are shown in Fig.~\ref{fig:southern_women_cycledist_gof_plots}, and
statnet goodness-of-fit plots in Fig.~\ref{fig:southern_women_gof}
(Appendix~\ref{sec:gof}).
Note that all four models fit acceptably well on all the statistics
included in the goodness-of-fit tests (degree distributions for each
mode, dyadwise shared partners, and geodesic distance distribution),
as well as the cycle length distributions. In particular, the fit to
four-cycle counts is good for all models, including Model~1 and Model~2,
which do not contain any terms to model four-cycles. It therefore
appears that Model~1 is the most parsimonious explanation of this
data, just as discussed in Wang~\etal{}~\cite[pp.~22]{wang09}.  This
model has a positive and statistically significant event two-star
parameter (b2star2), indicating ``greater discrepancies in the
popularity of events than expected in a random network''
\cite[pp.~22]{wang09}, taking into account the density (edges) and
actor two-star (b1star2) effects.

That ERG models containing only terms to model density and degree
distributions (and specifically not any terms to model dyadwise
shared partner distributions or four-cycles) also fit the four-cycle
count well indicates that the observed number of four-cycles could
have occurred simply by chance \cite[pp.~22]{wang09}. Note that the
same applies also to six-cycles. This is consistent with the results
for this network described by Opsahl \cite{opsahl13}, where the
observed value of the two-mode global clustering coefficient defined in that paper
is not extreme in the distribution of that coefficient in random networks.

\section{Conclusion}
\label{sec:conclusion}

The existing parameters for modelling shared partners or four-cycles
in ERGMs for two-mode networks frequently lead to convergence
problems, especially when parameters for both modes are included in
the model. A literature survey shows that the
majority of published ERG models of two-mode networks do not include
these parameters. In addition, the majority of published models
do not include an assessment goodness-of-fit to four-cycles.

In this work, we defined new ERGM effects to explicitly model
four-cycles, and in the case of two-mode networks, four-cycle counts
for the two modes separately. Simulation experiments show that ERGMs
using the new parameters are able to generate bipartite networks with
smoothly varying numbers of four-cycles, where the existing parametrisations
cannot.

These new parameters come with both conceptual and computational
costs. The conceptual cost is having to
use a more general dependence class in the dependence hierarchy,
without an underlying theoretical justification. The social
circuit dependence assumption has theoretical justifications, as does
partial inclusion ($\mathrm{PI}_1$) for pendant-triangle
configurations \cite{pattison24}, but we are forced to use the
more general class $\mathrm{D}_2$ simply by the dependency
induced by the new configuration, without any theoretical basis.

It is notable that our attempt to solve the problems with
near-degeneracy observed with the existing \kca{} and \kca{}
parameters with new parameters in the same dependence class
($\mathrm{I}_1$, ``social circuit'') failed due to these parameters
being even more prone to near-degeneracy
(Section~\ref{sec:altk4cycles}). These new parameters were motivated
by the observation (Section~\ref{sec:problems}) that the \kca{} and
\kcp{} statistics include not just four-cycles, but also simple
two-paths ($k$-two-paths with $k=1$), and that this could be a cause of
near-degeneracy with these parameters. However, as shown in
Section~\ref{sec:altk4cycles}, modifying the \kca{} and \kcp{}
parameters so that they only include four-cycles (and remain in the
same dependency class) did not solve the problem. The problem was only
solved (Section~\ref{sec:newstats}) by abandoning the dyadwise shared
partner structure of the \kca{} and \kcp{} statistics, and moving
instead to a node-oriented structure, counting the number of
four-cycles in which each node is involved (and using
``$\alpha$-inside'' weighting). This has the consequence that these
new statistics are no longer in the $\mathrm{I}_1$ dependence class,
but rather in the more general $\mathrm{D}_2$ class. Of course, just
because we were unable to solve the problems with the \kca{} and
\kcp{} parameters without creating a statistic in a more general
dependence class does not mean it is impossible to do so, and an
interesting direction for future work could be to either try to find a
statistic for four-cycles that is not prone to near-degeneracy and is
in the $\mathrm{I}_1$ (or perhaps $\mathrm{PI}_1$, $\mathrm{PI}_2$,
$\mathrm{I}_2$ or $\mathrm{D}_1$) dependence class, or to try to prove
that it is not possible.

The computational cost is incurred by 
having to traverse the two-neighbourhood of a dyad in computing the
change statistic, and even with precomputation of two-path counts 
\cite{stivala20}, this can be prohibitively slow (or require impractical amounts of memory) on dense and/or high
four-cycle count or very large networks, or those with very high
degree nodes.

Another shortcoming is the counterintuitive and possibly
confusing interpretation of the new parameters for bipartite
networks (BipartiteFourCyclesNodePowerA and
BipartiteFourCyclesNodePowerB; see
Section~\ref{sec:interpretation}). The new parameter for one-mode
networks (FourCyclesNodePower) is also applicable to two-mode
networks but does not treat the modes separately, and is relatively
straightforward to interpret simply as a four-cycle closure
parameter, so one solution is to use this, but this could omit
important differences in the modes in a two mode network. 
Even the relatively straightforward gwdegree parameter
used in statnet
causes significant confusion due to the counterintuitive
meaning of its sign
\cite{levy16poster,levy16,martin20,stivala20d}. Having to use such
relatively complex statistics, such as alternating, geometrically
weighted, or the ``$\alpha$-outside'' or ``$\alpha$-inside''
\cite{wilson17} weightings often solves the problem of near-degeneracy,
but comes at the cost of making the statistics difficult to
interpret \cite[p.~400]{blackburn23}. An
alternative solution to the conceptual (and possibly computational)
problem is to use instead the LOLOG model \cite{fellows18,clark22} or tapered
ERGM \cite{fellows17,blackburn23,ergm.tapered} where a simple
four-cycle parameter is unlikely to cause near-degeneracy problems,
much as these new models enable a simple triangle parameter in one-mode networks
where alternating or geometrically weighted parameters are
required in standard ERGMs
\cite{clark22,blackburn23,stivala23_bionet}. LOLOG, however, does not
(currently) handle bipartite networks \cite{clark22,lolog}, while tapered ERGM 
has the advantage of being
implemented in the statnet framework and can use any terms in the
statnet ergm package.

One further shortcoming of the new parameters is that the weighting
parameter $\alpha$ is fixed, and not estimated as part of the model.
A potential avenue of future work is to explore the possibility
of estimating this parameter in the context of a curved ERGM
in the statnet ergm package.

One final issue is fitting cycles of length larger than four in bipartite networks. Of particular importance are six-cycles, which have been suggested as the basis (rather than four-cycles) for measuring closure in two-mode networks \cite{opsahl13,vasquesfilho20,vasquesfilho20b}. The dependence class required for the new statistics described in this work also admits six-cycles as configurations, however we did not attempt to fit models with six-cycles as a parameter. It seems likely that, like the simple four-cycles parameter, attempting to do so would lead to problems with near-degeneracy, necessitating the creation of another weighted configuration analogous to those defined here for four-cycles (which would then be in another, even more general, dependence class).

\section*{Funding}

This work was supported by the Swiss National Science Foundation
[grant number 200778].

\section*{Acknowledgements}

This work was performed on the OzSTAR national facility at Swinburne
University of Technology. The OzSTAR program receives funding in part
from the Astronomy National Collaborative Research Infrastructure
Strategy (NCRIS) allocation provided by the Australian Government, and
from the Victorian Higher Education State Investment Fund (VHESIF)
provided by the Victorian Government.

\clearpage

\appendix
\setcounter{figure}{0}
\setcounter{table}{0}

\counterwithin{figure}{section}
\counterwithin{table}{section}
\renewcommand\thefigure{\thesection\arabic{figure}}
\renewcommand\thetable{\thesection\arabic{table}}

\section{Models contained in literature survey}
\label{sec:survey_models}

We used Google Scholar (search date 24 November 2022) to search for papers with terms ``bipartite`` and ``ergm'', ``bpnet'' and ``ergm'', ``statnet'' and ``bipartite'', and ``gwb1dsp'' or ``gwb2dsp'' (the latter being statnet \cite{handcock08,statnet} terms used for bipartite networks). We also included papers that we had prior knowledge of being about bipartite ERGMs, and followed backward and forward citations (using Google Scholar for the latter) to find other relevant papers. 

The resulting list of 117 models in 59 publications is shown in Table~\ref{tab:literature}. This table shows, for each network for which one or more models are presented in a paper, the size of both node sets in the network ($N_A$ is the number of nodes in node set $A$, and $N_B$ the number of nodes in node set $B$), the method used to estimate the model(s) in the paper, and the number of models for the network. Note that, in the case of multiple models for the same data, only a single model is counted; this is commonly the case for a model shown as developed incrementally, from a simple baseline model, and including more effects in subsequent models. In such cases, we consider only the final model. Multiple models are counted for cases such as models estimated for the same network at different time points, or for the same node set but with different edge types.

In Table~\ref{tab:literature_number_estimated_bpnet} of the main text,
the counts for the $k$-two-path parameters \kcp{} and \kca{} count
those estimated with \kcp{} and \kca{} in BPNet, and XACA and XACB
(respectively) in MPNet.

In Table~\ref{tab:literature_number_estimated_statnet} of the main text,
the counts for gwb1dsp and
gwb2dsp include those for the bipartite geometrically weighted
non-edgewise shared partner terms gwb1nsp and gwb2nsp, which are
equivalent to gwb1dsp and gwb2dsp, respectively, for bipartite
networks.  

\begin{longtable}{lp{0.30\linewidth}rrlr}
  \caption{Models included in literature survey of bipartite ERGM applications} \\
      \hline
      &       & \multicolumn{2}{c}{Network size} & Estim. & Num.  \\
      Citation & Network description & $N_A$ & $N_B$ & method &  models \\
      \hline
      \endhead
  \label{tab:literature}%
\cite{agneessens04} & Theatregoers and theatre performances & 290   & 24    & MPLE  & 1 \\
\cite{agneessens08} & Theatregoers and theatre performances & 290   & 24    & MPLE & 1 \\
\cite{balest19} & Municipalities and public utilities & 116   & 120   & Bergm  & 1 \\
\cite{benton17} & Shareholder activists and firms & 162   & 220   & statnet  &  1\\
\cite{berardo14} & Organizations and projects & 198   & 95    & BPNet  & 4 \\
\cite{bi21}  & Customers and products (cars) & 5000  & 250   & statnet  & 6 \\
\cite{bond12} & Affiliation network (club membership) & 257   & 15    & BPNet  & 1 \\
\cite{brandenberger15} & Actors and issues in water policy & ? & ? & statnet  &  1\\
\cite{conaldi13} & Software contributors and software bugs & 72    & 737   & BPNet  & 1 \\
\cite{desisto22} & Lemur species and plant genera & 55    & 590   & statnet  & 1 \\
\cite{duxbury18} & Buyers and sellers in an online drug cryptomarket & 706   & 57    & statnet  & 1 \\
\cite{duxbury18c} & Buyers and sellers in an online drug cryptomarket & 706   & 57    & statnet  & 1 \\
\cite{faust02b} & Co-attendance of Soviet politicians at events & 67    & 1816  & MPLE  & 8 \\
\cite{fried22} & Climate change adaptation actors and issues & 659   & 19    & statnet  & 1 \\
\cite{fritz23} & Patents and inventors (temporal) & 78\,412 & 126\,388 & statnet  & 1 \\
\cite{fuzessy22} & Fruit-frugivore interactions (vertebrates and plants), Continuum & 133   & 315   & statnet  & 1 \\
\cite{fuzessy22} & Fruit-frugivore interactions (vertebrates and plants), Fragment & 54    & 58    & statnet  & 1 \\
\cite{gallemore15} & Provincial and national organizations & 29    & 52    & statnet  & 1 \\
\cite{gondal11} & Authors and papers & 2\,200  & 76    & BPNet  & 1\\
\cite{gondal18} & University departments and specializations & 101   & 77    & MPNet  & 1 \\
\cite{hamilton18} & Policy actors and decision-making forums  & 109   & 84    & statnet  & 1 \\
\cite{harrigan13_in_lusher13book} & Director interlock, largest 248 corporations by revenue in Australia & 1\,251  & 248   & BPNet  & 1 \\
\cite{hazir12} & Organizations and research projects & 1\,316  & 237   & BPNet  & 2 \\
\cite{heaney18_preprint} & Interest groups and lobbying coalitions & 171   & 74    & statnet  & 1 \\
\cite{huang09} & Characters and their groups in an online game & 465   & 396   & BPNet  & 7 \\
\cite{jasny12} & Organizations and forums in watershed policy & ? & ? & statnet  & 1 \\
\cite{jasny15} & Organizations and policy institutions & 527   & 146   & BPNet  & 1 \\
\cite{keegan12} & Wikipedia articles and editors & 14\,292 & 249   & statnet  & 1 \\
\cite{kevork22} & Southern Women & 18    & 14    & statnet  & 1 \\
\cite{kevork22} & World city network (global firms and cities) & 100   & 315   & statnet  & 1 \\
\cite{kevork22} & Inventors and patents & 10\,251 & 8\,206  & statnet  &1  \\
\cite{khalilzadeh18} & Travel destination countries and attitudes towards them & 47    & 22    & statnet  & 1 \\
\cite{lai17} & Disaster response organizations and resource types & 27    & 10    & BPNet  & 3 \\
\cite{lai17} & Affected neighbourhoods and resource contacts & 19    & 10    & BPNet  & 4 \\
\cite{leifeld14_preprint} & EU parliamentary chambers and proposals & 39    & 650   & statnet  & 1 \\
\cite{li21b} & Collaborations for hazard mitigation & 95    & 198   & statnet  & 1 \\
\cite{lu22}  & Online doctor-patient consultations & ? & ? &  statnet & 1 \\
\cite{lubell14} & Actors and institutions in water management & 167   & 220   & BPNet  & 2 \\
\cite{lubell22} & Actors and forums in adapting to sea-level rise & 82    & 9     & statnet  & 1 \\
\cite{lubell22} & Actors and forums in adapting to sea-level rise & 647   & 103   & statnet  & 1 \\
\cite{margolin12} & Voluntary collaborative project teams & 170   & 124   & BPNet  & 1 \\
\cite{mauldin21} & Aged care residents and group activities & 35    & 563   & statnet  & 1 \\
\cite{mcallister14} & Organizations and policy forums & 52    & 16    & BPNet  & 2 \\
\cite{mcallister15} & Stakeholders and committees/working groups & 152   & 9     & BPNet  & 2 \\
\cite{mcallister15} & Stakeholders and committees/working groups & 180   & 9     & BPNet  & 5 \\
\cite{mcallister15b} & Stakeholders and forums in urban development & 60    & 14    & MPNet  & 1 \\
\cite{mcallister17} & Stakeholders and forums in myrtle rust response & 259   & 12    & MPNet  & 1 \\
\cite{metz18} & Actors and instruments in water policy & 31    & 15    & statnet  & 1 \\
\cite{niekamp13} & "Swinging" couples and  venues & 57    & 39    & BPNet  & 1 \\
\cite{nita16} & Organizations and projects in EU conservation (UK) & 120   & 46    & BPNet  & 1 \\
\cite{nita16} & Organizations and projects in EU conservation (NL) & 54    & 40    & BPNet  & 1 \\
\cite{nita16} & Organizations and projects in EU conservation (PT) & 156   & 63    & BPNet  & 1 \\
\cite{nita16} & Organizations and projects in EU conservation (GR) & 124   & 57    & BPNet  & 1 \\
\cite{nita16} & Organizations and projects in EU conservation (RO) & 119   & 48    & BPNet  & 1 \\
\cite{nita16} & Organizations and projects in EU conservation (LV) & 122   & 28    & BPNet  & 1 \\
\cite{norbutas18} & Buyers and sellers in an online drug cryptomarket & 3\,542  & 463   & statnet  & 1 \\
\cite{park22} & Citations from web top-level domains to health agency  websites & ? & 148   & statnet  & 1 \\
\cite{ren23} & Stakeholders and resilience planning documents & ? & 39    & statnet  & 1 \\
\cite{rocha15} & Drivers and regime shifts in social-ecological systems & 57    & 25    & statnet  & 1 \\
\cite{scott17} & Actors and forums in collaborative governance regimes & 400   & 57    & statnet  & 3 \\
\cite{sha19} & Customers and car models & 5\,000  & 281   & statnet  & 1 \\
\cite{sha19b} & Participants in design crowdsourcing contests & 3\,462  & 96    & statnet  & 1 \\
\cite{shin22} & Deforestation emissions reduction projects and countries & 480   & 57    & MPNet  & 1 \\
\cite{shjarback18} & Crime control bills and their sponsors & 1\,304  & 221   & statnet  & 1 \\
\cite{stephens16} & Employees and ideas in intraorganizational crowdsourcing (stage 1) & 213   & 236   & BPNet  & 3 \\
\cite{stephens16} & Employees and ideas in intraorganizational crowdsourcing (stage 2) & 685   & 578   & BPNet  & 3 \\
\cite{stephens16} & Employees and ideas in intraorganizational crowdsourcing (full) & 768   & 640   & BPNet  & 3 \\
\cite{wang09} & Southern Women & 18    & 14    & BPNet  & 1 \\
\cite{wang09} & Director interlock & 366   & 50    & BPNet  & 1 \\
\cite{wang09} & Director interlock & 255   & 198   & BPNet  &  1\\
\cite{wang13_bipartite} & Student activists youth leaders by event & 14    & 49    & BPNet  & 1 \\
\cite{wang13_bipartite} & Student activists organization by event & 23    & 49    & BPNet  & 1 \\
\cite{wang20b} & Participants and teams in online crowdsourcing & 2\,100  & 946   & statnet  & 1 \\
\cite{wonka19} & EU Parliamentarians and information sources & 77    & 18    & statnet  & 1 \\
\cite{zhu13} & Individuals and teams in an online game & 333   & 426   & BPNet  & 1 \\
\hline
\end{longtable}

\clearpage

\section{Goodness-of-fit plots}
\label{sec:gof}

\begin{figure}[h!]
  \centering
  \includegraphics{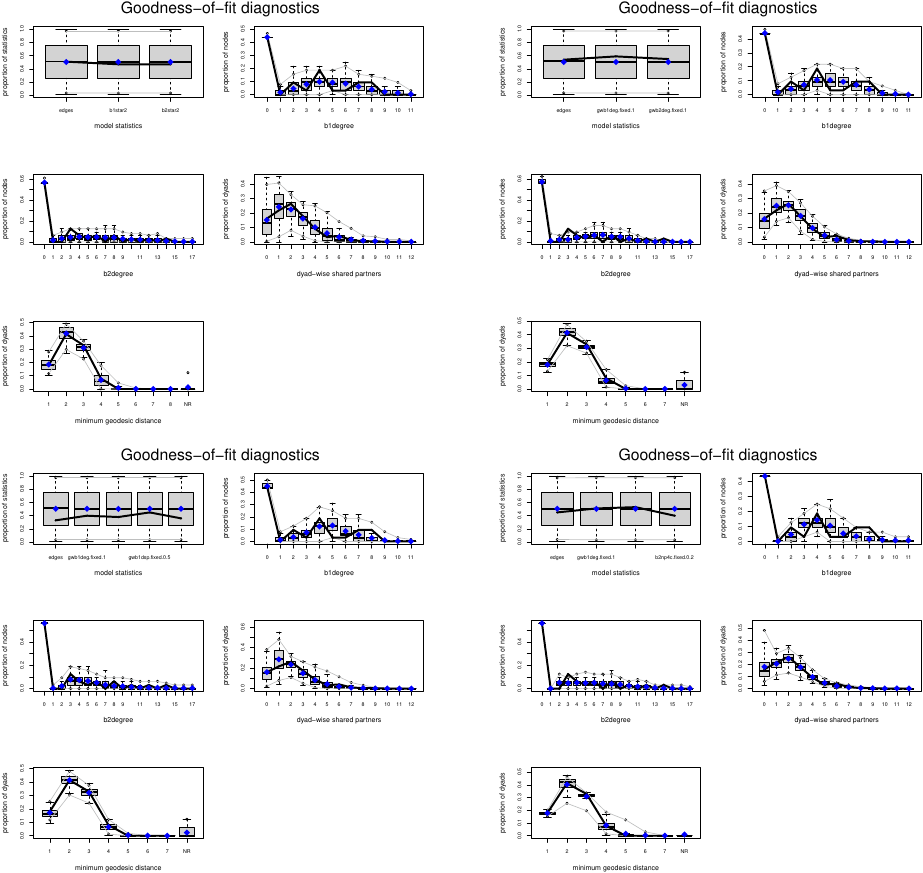}
  \caption{Statnet goodness-of-fit plots for the Southern Women ERGM
    (Table~\ref{tab:southern_women_ergm_results}) Model 1 (top left), Model 2
    (top right), Model 3 (bottom left), and Model 4 (bottom right).}
  \label{fig:southern_women_gof}
\end{figure}

Goodness-of-fit plots for the Southern Women network ERG models in
Section~\ref{sec:statnet_example} are shown in
Fig.~\ref{fig:southern_women_gof}.

\clearpage


\begin{thebibliography}{100}

\bibitem{breiger74}
Ronald~L Breiger.
\newblock The duality of persons and groups.
\newblock {\em Soc Forces}, 53(2):181--190, 1974.

\bibitem{neal24c}
Zachary~P. Neal, Annabell Cadieux, Diego Garlaschelli, Nicholas~J. Gotelli,
  Fabio Saracco, Tiziano Squartini, Shade~T. Shutters, Werner Ulrich, Guanyang
  Wang, and Giovanni Strona.
\newblock Pattern detection in bipartite networks: A review of terminology,
  applications, and methods.
\newblock {\em PLOS Complex Syst}, 1(2):e0000010, 10 2024.

\bibitem{wang13_multilevel}
Peng Wang, Garry Robins, Philippa Pattison, and Emmanuel Lazega.
\newblock Exponential random graph models for multilevel networks.
\newblock {\em Soc Netw}, 35(1):96--115, 2013.

\bibitem{latapy08}
Matthieu Latapy, Cl{\'e}mence Magnien, and Nathalie Del~Vecchio.
\newblock Basic notions for the analysis of large two-mode networks.
\newblock {\em Soc Netw}, 30(1):31--48, 2008.

\bibitem{everett13}
Martin~G Everett and Stephen~P Borgatti.
\newblock The dual-projection approach for two-mode networks.
\newblock {\em Soc Netw}, 35(2):204--210, 2013.

\bibitem{everett16}
Martin~G Everett.
\newblock Centrality and the dual-projection approach for two-mode social
  network data.
\newblock {\em Method Innov}, 9:2059799116630662, 2016.

\bibitem{granovetter73}
Mark~S. Granovetter.
\newblock The strength of weak ties.
\newblock {\em Am J Sociol}, 78(6):1360--1380, 1973.

\bibitem{mathworld_bipartite_graph}
Eric~W. Weisstein.
\newblock Bipartite graph. {From MathWorld---A Wolfram Web Resource}.
\newblock \url{https://mathworld.wolfram.com/BipartiteGraph.html}, 2024.

\bibitem{robins04b}
Garry Robins and Malcolm Alexander.
\newblock Small worlds among interlocking directors: Network structure and
  distance in bipartite graphs.
\newblock {\em Comput Math Organ Theory}, 10(1):69--94, 2004.

\bibitem{opsahl13}
Tore Opsahl.
\newblock Triadic closure in two-mode networks: Redefining the global and local
  clustering coefficients.
\newblock {\em Soc Netw}, 35(2):159--167, 2013.

\bibitem{koskinen12}
Johan Koskinen and Christofer Edling.
\newblock Modelling the evolution of a bipartite network—peer referral in
  interlocking directorates.
\newblock {\em Soc Netw}, 34(3):309--322, 2012.

\bibitem{vasquesfilho20}
Demival Vasques~Filho and Dion~RJ O'Neale.
\newblock Transitivity and degree assortativity explained: The bipartite
  structure of social networks.
\newblock {\em Phys Rev E}, 101(5):052305, 2020.

\bibitem{vasquesfilho20b}
Demival Vasques~Filho and Dion~RJ O'Neale.
\newblock The role of bipartite structure in {R\&D} collaboration networks.
\newblock {\em J Complex Netw}, 8(4):cnaa016, 10 2020.

\bibitem{lusher13}
Dean Lusher, Johan Koskinen, and Garry Robins, editors.
\newblock {\em Exponential Random Graph Models for Social Networks: Theory,
  Methods, and Applications}.
\newblock Structural Analysis in the Social Sciences. Cambridge University
  Press, New York, 2013.

\bibitem{amati18}
Viviana Amati, Alessandro Lomi, and Antonietta Mira.
\newblock Social network modeling.
\newblock {\em Annu Rev Stat Appl}, 5:343--369, 2018.

\bibitem{koskinen20}
J.~Koskinen.
\newblock Exponential random graph modelling.
\newblock In P.~Atkinson, S.~Delamont, A.~Cernat, J.W. Sakshaug, and R.A.
  Williams, editors, {\em SAGE Research Methods Foundations}. SAGE, London,
  2020.
\newblock \url{https://doi.org/10.4135/9781526421036888175}.

\bibitem{koskinen23}
Johan Koskinen.
\newblock Exponential random graph models.
\newblock In John McLevey, John Scott, and Peter~J Carrington, editors, {\em
  The Sage Handbook of Social Network Analysis}, chapter~33. Sage, second
  edition, 2023.

\bibitem{cimini19}
Giulio Cimini, Tiziano Squartini, Fabio Saracco, Diego Garlaschelli, Andrea
  Gabrielli, and Guido Caldarelli.
\newblock The statistical physics of real-world networks.
\newblock {\em Nat Rev Phys}, 1(1):58--71, 2019.

\bibitem{ghafouri20}
Saeid Ghafouri and Seyed~Hossein Khasteh.
\newblock A survey on exponential random graph models: an application
  perspective.
\newblock {\em PeerJ Comput Sci}, 6:e269, 2020.

\bibitem{giacomarra23}
Francesco Giacomarra, Gianmarco Bet, and Alessandro Zocca.
\newblock Generating synthetic power grids using exponential random graph
  models.
\newblock {\em PRX Energy}, 3(2):023005, 2024.

\bibitem{wang09}
Peng Wang, Ken Sharpe, Garry~L Robins, and Philippa~E Pattison.
\newblock Exponential random graph (p*) models for affiliation networks.
\newblock {\em Soc Netw}, 31(1):12--25, 2009.

\bibitem{wang13_bipartite}
Peng Wang, Philippa Pattison, and Garry Robins.
\newblock Exponential random graph model specifications for bipartite
  networks—a dependence hierarchy.
\newblock {\em Soc Netw}, 35(2):211--222, 2013.

\bibitem{wang13_in_lusher13book}
Peng Wang.
\newblock Exponential random graph model extensions: Models for multiple
  networks and bipartite networks.
\newblock In Dean Lusher, Johan Koskinen, and Garry Robins, editors, {\em
  Exponential Random Graph Models for Social Networks: Theory, Methods, and
  Applications}, chapter~10, pages 115--129. Cambridge University Press, New
  York, 2013.

\bibitem{bomiriya23}
Rashmi~P Bomiriya, Alina~R Kuvelkar, David~R Hunter, and Steffen Triebel.
\newblock Modeling homophily in exponential-family random graph models for
  bipartite networks.
\newblock {\em arXiv preprint arXiv:2312.05673v1}, 2023.

\bibitem{pattison13_in_lusher13book}
Philippa~E. Pattison and T.A.B. Snijders.
\newblock Modeling social networks: Next steps.
\newblock In Dean Lusher, Johan Koskinen, and Garry Robins, editors, {\em
  Exponential Random Graph Models for Social Networks: Theory, Methods, and
  Applications}, chapter~22, pages 287--301. Cambridge University Press, New
  York, 2013.

\bibitem{geyer92}
Charles~J Geyer and Elizabeth~A Thompson.
\newblock Constrained {Monte Carlo} maximum likelihood for dependent data.
\newblock {\em J R Stat Soc B}, 54(3):657--683, 1992.

\bibitem{snijders02}
Tom A.~B. Snijders.
\newblock Markov chain {Monte Carlo} estimation of exponential random graph
  models.
\newblock {\em J Soc Struct}, 3(2):1--40, 2002.

\bibitem{hunter12}
David~R Hunter, Pavel~N Krivitsky, and Michael Schweinberger.
\newblock Computational statistical methods for social network models.
\newblock {\em J Comput Graph Stat}, 21(4):856--882, 2012.

\bibitem{byshkin16}
Maksym Byshkin, Alex Stivala, Antonietta Mira, Rolf Krause, Garry Robins, and
  Alessandro Lomi.
\newblock Auxiliary parameter {MCMC} for exponential random graph models.
\newblock {\em J Stat Phys}, 165(4):740--754, 2016.

\bibitem{byshkin18}
Maksym Byshkin, Alex Stivala, Antonietta Mira, Garry Robins, and Alessandro
  Lomi.
\newblock Fast maximum likelihood estimation via equilibrium expectation for
  large network data.
\newblock {\em Sci Rep}, 8(1):11509, 2018.

\bibitem{borisenko20}
Alexander Borisenko, Maksym Byshkin, and Alessandro Lomi.
\newblock A simple algorithm for scalable {Monte Carlo} inference.
\newblock {\em arXiv preprint arXiv:1901.00533v4}, 2020.

\bibitem{handcock03}
Mark~S Handcock.
\newblock Assessing degeneracy in statistical models of social networks.
\newblock Working Paper no. 39, Center for Statistics and the Social Sciences,
  University of Washington, 2003.
\newblock \url{https://csss.uw.edu/Papers/wp39.pdf}.

\bibitem{snijders06}
Tom A.~B. Snijders, Philippa~E. Pattison, Garry~L. Robins, and Mark~S.
  Handcock.
\newblock New specifications for exponential random graph models.
\newblock {\em Sociol Methodol}, 36(1):99--153, 2006.

\bibitem{fellows17}
Ian Fellows and Mark Handcock.
\newblock Removing phase transitions from {Gibbs} measures.
\newblock In Aarti Singh and Jerry Zhu, editors, {\em Proceedings of the 20th
  International Conference on Artificial Intelligence and Statistics},
  volume~54 of {\em Proceedings of Machine Learning Research}, pages 289--297,
  20--22 Apr 2017.

\bibitem{schweinberger11}
Michael Schweinberger.
\newblock Instability, sensitivity, and degeneracy of discrete exponential
  families.
\newblock {\em J Am Stat Assoc}, 106(496):1361--1370, 2011.

\bibitem{chatterjee13}
Sourav Chatterjee and Persi Diaconis.
\newblock Estimating and understanding exponential random graph models.
\newblock {\em Ann Stat}, 41(5):2428--2461, 2013.

\bibitem{schweinberger20}
Michael Schweinberger.
\newblock Consistent structure estimation of exponential-family random graph
  models with block structure.
\newblock {\em Bernoulli}, 26(2):1205--1233, 2020.

\bibitem{blackburn23}
Bart Blackburn and Mark~S Handcock.
\newblock Practical network modeling via tapered exponential-family random
  graph models.
\newblock {\em J Comput Graph Stat}, 32(2):388--401, 2023.

\bibitem{robins07}
Garry Robins, Tom A.~B. Snijders, Peng Wang, Mark Handcock, and Philippa
  Pattison.
\newblock Recent developments in exponential random graph ($p^*$) models for
  social networks.
\newblock {\em Soc Netw}, 29(2):192--215, 2007.

\bibitem{koskinen13}
J.~Koskinen and G.~Daraganova.
\newblock Exponential random graph model fundamentals.
\newblock In Dean Lusher, Johan Koskinen, and Garry Robins, editors, {\em
  Exponential Random Graph Models for Social Networks: Theory, Methods, and
  Applications}, chapter~6, pages 49--76. Cambridge University Press, New York,
  2013.

\bibitem{hunter07}
David~R. Hunter.
\newblock Curved exponential family models for social networks.
\newblock {\em Soc Netw}, 29(2):216--230, 2007.

\bibitem{stivala23_alaam}
Alex Stivala.
\newblock Overcoming near-degeneracy in the autologistic actor attribute model.
\newblock {\em arXiv preprint arXiv:2309.07338v2}, 2023.

\bibitem{hunter06}
David~R. Hunter and Mark~S. Handcock.
\newblock Inference in curved exponential family models for networks.
\newblock {\em J Comput Graph Stat}, 15(3):565--583, 2006.

\bibitem{mpnet14}
Peng Wang, Garry Robins, Philippa Pattison, and JH~Koskinen.
\newblock {\em {MPNet}: Program for the simulation and estimation of (p*)
  exponential random graph models for multilevel networks}.
\newblock Melbourne School of Psychological Sciences, The University of
  Melbourne, 2014.
\newblock \url{http://www.melnet.org.au/s/MPNetManual.pdf}.

\bibitem{mpnet22}
Peng Wang, Alex Stivala, Garry Robins, Philippa Pattison, Johan Koskinen, and
  Alessandro Lomi.
\newblock {\em PNet: Program for the simulation and estimation of (p*)
  exponential random graph models for multilevel networks}, 2022.
\newblock \url{http://www.melnet.org.au/s/MPNetManual2022.pdf}.

\bibitem{bpnet}
P.~Wang, G.~Robins, and P.~Pattison.
\newblock {\em PNet: program for the estimation and simulation of $p^*$
  exponential random graph models}.
\newblock Department of Psychology, The University of Melbourne, 2009.
\newblock \url{https://www.melnet.org.au/s/PNetManual.pdf}.

\bibitem{handcock08}
Mark~S. Handcock, David~R. Hunter, Carter~T. Butts, Steven~M. Goodreau, and
  Martina Morris.
\newblock statnet: Software tools for the representation, visualization,
  analysis and simulation of network data.
\newblock {\em J Stat Softw}, 24(1):1--11, 2008.

\bibitem{hunter2008ergm}
David~R Hunter, Mark~S Handcock, Carter~T Butts, Steven~M Goodreau, and Martina
  Morris.
\newblock ergm: A package to fit, simulate and diagnose exponential-family
  models for networks.
\newblock {\em J Stat Softw}, 24(3):1--29, 2008.

\bibitem{hummel12}
Ruth~M. Hummel, David~R. Hunter, and Mark~S. Handcock.
\newblock Improving simulation-based algorithms for fitting {ERGMs}.
\newblock {\em J Comput Graph Stat}, 21(4):920--939, 2012.

\bibitem{statnet}
Mark~S. Handcock, David~R. Hunter, Carter~T. Butts, Steven~M. Goodreau,
  Pavel~N. Krivitsky, Skye Bender-deMoll, and Martina Morris.
\newblock {\em statnet: Software Tools for the Statistical Analysis of Network
  Data}.
\newblock The Statnet Project (\url{http://www.statnet.org}), 2016.
\newblock R package version 2019.6.
  \url{https://CRAN.R-project.org/package=statnet}.

\bibitem{ergm}
Mark~S. Handcock, David~R. Hunter, Carter~T. Butts, Steven~M. Goodreau,
  Pavel~N. Krivitsky, and Martina Morris.
\newblock {\em ergm: Fit, Simulate and Diagnose Exponential-Family Models for
  Networks}.
\newblock The Statnet Project (\url{https://statnet.org}), 2024.
\newblock R package version 4.7.5.
  \url{https://CRAN.R-project.org/package=ergm}.

\bibitem{ergm4}
Pavel~N. Krivitsky, David~R. Hunter, Martina Morris, and Chad Klumb.
\newblock ergm 4: New features for analyzing exponential-family random graph
  models.
\newblock {\em J Stat Softw}, 105(6):1–44, 2023.

\bibitem{ergm4_computational}
Pavel~N Krivitsky, David~R Hunter, Martina Morris, and Chad Klumb.
\newblock ergm 4: Computational improvements.
\newblock {\em arXiv preprint arXiv:2203.08198v1}, 2022.

\bibitem{caimo14}
Alberto Caimo and Nial Friel.
\newblock {Bergm}: {B}ayesian exponential random graphs in {R}.
\newblock {\em J Stat Softw}, 61(2):1--25, 2014.

\bibitem{bergm}
Alberto Caimo, Lampros Bouranis, Robert Krause, and Nial Friel.
\newblock Statistical network analysis with {Bergm}.
\newblock {\em J Stat Softw}, 104(1):1--23, 2022.

\bibitem{balest19}
Jessica Balest, Laura Secco, Elena Pisani, and Alberto Caimo.
\newblock Sustainable energy governance in {South Tyrol} ({Italy}): A
  probabilistic bipartite network model.
\newblock {\em J Clean Prod}, 221:854--862, 2019.

\bibitem{agneessens08}
Filip Agneessens and Henk Roose.
\newblock Local structural properties and attribute characteristics in 2-mode
  networks: p* models to map choices of theater events.
\newblock {\em J Math Sociol}, 32(3):204--237, 2008.

\bibitem{benton17}
Richard~A. Benton and Jihae You.
\newblock Endogenous dynamics in contentious fields: Evidence from the
  shareholder activism network, 2006–2013.
\newblock {\em Socius}, 3:2378023117705231, 2017.

\bibitem{lubell22}
Mark Lubell and Matthew Robbins.
\newblock Adapting to sea-level rise: Centralization or decentralization in
  polycentric governance systems?
\newblock {\em Policy Stud J}, 50(1):143--175, 2022.

\bibitem{watts98}
Duncan~J. Watts and Steven~H. Strogatz.
\newblock Collective dynamics of `small-world' networks.
\newblock {\em Nature}, 393(6684):440--442, 1998.

\bibitem{newman03c}
Mark~EJ Newman and Juyong Park.
\newblock Why social networks are different from other types of networks.
\newblock {\em Phys Rev E}, 68(3):036122, 2003.

\bibitem{clark22}
Duncan~A Clark and Mark~S Handcock.
\newblock Comparing the real-world performance of exponential-family random
  graph models and latent order logistic models for social network analysis.
\newblock {\em J R Stat Soc A}, 185(2):566--587, 2022.

\bibitem{fellows18}
Ian~E Fellows.
\newblock A new generative statistical model for graphs: The latent order
  logistic ({LOLOG}) model.
\newblock {\em arXiv preprint arXiv:1804.04583v1}, 2018.

\bibitem{pauksztat11}
Birgit Pauksztat, Christian Steglich, and Rafael Wittek.
\newblock Who speaks up to whom? a relational approach to employee voice.
\newblock {\em Soc Netw}, 33(4):303--316, 2011.

\bibitem{doreian12}
Patrick Doreian and Norman Conti.
\newblock Social context, spatial structure and social network structure.
\newblock {\em Soc Netw}, 34(1):32--46, 2012.
\newblock Capturing Context: Integrating Spatial and Social Network Analyses.

\bibitem{wong15}
Ling Heng~Henry Wong, André~F. Gygax, and Peng Wang.
\newblock Board interlocking network and the design of executive compensation
  packages.
\newblock {\em Soc Netw}, 41:85--100, 2015.

\bibitem{goodreau07}
Steven~M. Goodreau.
\newblock Advances in exponential random graph ($p^*$) models applied to a
  large social network.
\newblock {\em Soc Netw}, 29(2):231--248, 2007.

\bibitem{heidler14}
Richard Heidler, Markus Gamper, Andreas Herz, and Florian E{\ss}er.
\newblock Relationship patterns in the 19th century: The friendship network in
  a {German} boys’ school class from 1880 to 1881 revisited.
\newblock {\em Soc Netw}, 37:1--13, 2014.

\bibitem{sailer12}
Kerstin Sailer and Ian McCulloh.
\newblock Social networks and spatial configuration—how office layouts drive
  social interaction.
\newblock {\em Soc Netw}, 34(1):47--58, 2012.
\newblock Capturing Context: Integrating Spatial and Social Network Analyses.

\bibitem{fischer15}
Manuel Fischer and Pascal Sciarini.
\newblock Unpacking reputational power: Intended and unintended determinants of
  the assessment of actors’ power.
\newblock {\em Soc Netw}, 42:60--71, 2015.

\bibitem{toivonen09}
Riitta Toivonen, Lauri Kovanen, Mikko Kivelä, Jukka-Pekka Onnela, Jari
  Saramäki, and Kimmo Kaski.
\newblock A comparative study of social network models: Network evolution
  models and nodal attribute models.
\newblock {\em Soc Netw}, 31(4):240--254, 2009.

\bibitem{ackland111}
Robert Ackland and Mathieu O’Neil.
\newblock Online collective identity: The case of the environmental movement.
\newblock {\em Soc Netw}, 33(3):177--190, 2011.

\bibitem{anderson99b}
Carolyn~J Anderson, Stanley Wasserman, and Bradley Crouch.
\newblock A p* primer: logit models for social networks.
\newblock {\em Soc Netw}, 21(1):37--66, 1999.

\bibitem{stivala20}
Alex Stivala, Garry Robins, and Alessandro Lomi.
\newblock Exponential random graph model parameter estimation for very large
  directed networks.
\newblock {\em PLoS One}, 15(1):e0227804, 2020.

\bibitem{harrigan13_in_lusher13book}
Nicholas Harrigan and Matthew Bond.
\newblock Differential impact of directors' social and financial capital on
  corporate interlock formation.
\newblock In Dean Lusher, Johan Koskinen, and Garry Robins, editors, {\em
  Exponential Random Graph Models for Social Networks: Theory, Methods, and
  Applications}, chapter~20, pages 260--271. Cambridge University Press, New
  York, 2013.

\bibitem{stivala20c}
Alex Stivala.
\newblock Geodesic cycle length distributions in delusional and other social
  networks.
\newblock {\em J Soc Struct}, 21(1):35--76, 2020.

\bibitem{stivala23_geodesic}
Alex Stivala.
\newblock Geodesic cycle length distributions in fictional character networks.
\newblock {\em arXiv preprint arXiv:2303.11597v1}, 2023.

\bibitem{martin20}
John~Levi Martin.
\newblock Comment on geodesic cycle length distributions in delusional and
  other social networks.
\newblock {\em J Soc Struct}, 21(1):77--93, 2020.

\bibitem{wilson17}
James~D Wilson, Matthew~J Denny, Shankar Bhamidi, Skyler~J Cranmer, and Bruce~A
  Desmarais.
\newblock Stochastic weighted graphs: Flexible model specification and
  simulation.
\newblock {\em Soc Netw}, 49:37--47, 2017.

\bibitem{pattison24}
Philippa~E. Pattison, Garry~L. Robins, Tom~A.B. Snijders, and Peng Wang.
\newblock Exponential random graph models and pendant-triangle statistics.
\newblock {\em Soc Netw}, 79:187--197, 2024.

\bibitem{pattison02}
Philippa Pattison and Garry Robins.
\newblock Neighborhood--based models for social networks.
\newblock {\em Sociol Methodol}, 32(1):301--337, 2002.

\bibitem{pattison04}
Philippa Pattison and Garry Robins.
\newblock Building models for social space: Neighourhood-based models for
  social networks and affiliation structures.
\newblock {\em Math \& Sci Hum}, 42(168):11--29, 2004.

\bibitem{hunter13}
David~R. Hunter, Steven~M. Goodreau, and Mark~S. Handcock.
\newblock ergm.userterms: A template package for extending statnet.
\newblock {\em J Stat Softw}, 52(2):1–25, 2013.

\bibitem{hunter19}
David~R Hunter and Steven~M Goodreau.
\newblock Extending {ERGM} functionality within statnet: Building custom user
  terms.
\newblock
  \url{https://statnet.org/workshop-ergm-userterms/ergm.userterms_tutorial.pdf},
  2019.
\newblock Statnet Development Team.

\bibitem{uno14}
Takeaki Uno and Hiroko Satoh.
\newblock An efficient algorithm for enumerating chordless cycles and chordless
  paths.
\newblock In Sa{\v{s}}o D{\v{z}}eroski, Pan{\v{c}}e Panov, Dragi Kocev, and
  Ljup{\v{c}}o Todorovski, editors, {\em International Conference on Discovery
  Science}, volume 8777 of {\em LNAI}, pages 313--324. Springer, 2014.

\bibitem{R-manual}
{R Core Team}.
\newblock {\em R: A Language and Environment for Statistical Computing}.
\newblock R Foundation for Statistical Computing, Vienna, Austria, 2022.

\bibitem{csardi06}
G\'abor Cs\'ardi and Tamas Nepusz.
\newblock The igraph software package for complex network research.
\newblock {\em InterJournal}, Complex Systems:1695, 2006.

\bibitem{antonov23}
Michael Antonov, G{\'a}bor Cs{\'a}rdi, Szabolcs Horv{\'a}t, Kirill M{\"u}ller,
  Tam{\'a}s Nepusz, Daniel Noom, Ma{\"e}lle Salmon, Vincent Traag,
  Brooke~Foucault Welles, and Fabio Zanini.
\newblock igraph enables fast and robust network analysis across programming
  languages.
\newblock {\em arXiv preprint arXiv:2311.10260v1}, 2023.

\bibitem{ggplot2}
Hadley Wickham.
\newblock {\em ggplot2: Elegant Graphics for Data Analysis}.
\newblock Springer-Verlag, New York, 2016.

\bibitem{morris08}
Martina Morris, Mark Handcock, and David Hunter.
\newblock Specification of exponential-family random graph models: Terms and
  computational aspects.
\newblock {\em J Stat Softw}, 24(4):1--24, 2008.

\bibitem{davis41}
Allison Davis, Burleigh~B Gardner, Mary~R Gardner, and W~Lloyd Warner.
\newblock {\em Deep South: A sociological anthropological study of caste and
  class}.
\newblock University of Chicago Press, 1941.

\bibitem{krivitsky08}
Pavel~N. Krivitsky and Mark~S. Handcock.
\newblock Fitting position latent cluster models for social networks with
  latentnet.
\newblock {\em J Stat Softw}, 24(5), 2008.

\bibitem{latentnet}
Pavel~N. Krivitsky and Mark~S. Handcock.
\newblock {\em latentnet: Latent Position and Cluster Models for Statistical
  Networks}.
\newblock The Statnet Project (\url{https://statnet.org}), 2024.
\newblock R package version 2.11.0.
  \url{https://CRAN.R-project.org/package=latentnet}.

\bibitem{lerner22}
Jürgen Lerner and Alessandro Lomi.
\newblock A dynamic model for the mutual constitution of individuals and
  events.
\newblock {\em J Complex Netw}, 10(2):cnac004, 03 2022.

\bibitem{kevork22}
Sevag Kevork and Göran Kauermann.
\newblock Bipartite exponential random graph models with nodal random effects.
\newblock {\em Soc Netw}, 70:90--99, 2022.

\bibitem{everett18}
Martin~G Everett, Chiara Broccatelli, Stephen~P Borgatti, and Johan Koskinen.
\newblock Measuring knowledge and experience in two mode temporal networks.
\newblock {\em Soc Netw}, 55:63--73, 2018.

\bibitem{leifeld13}
Philip Leifeld.
\newblock {texreg}: Conversion of statistical model output in {R} to {\LaTeX}
  and {HTML} tables.
\newblock {\em J Stat Softw}, 55(8):1--24, 2013.

\bibitem{levy16poster}
Michael Levy, Mark Lubell, Philip Leifeld, and Skyler Cranmer.
\newblock Interpretation of gw-degree estimates in {ERGM}s, June 2016.
\newblock \url{https://doi.org/10.6084/m9.figshare.3465020.v1}.

\bibitem{levy16}
Michael Levy.
\newblock gwdegree: Improving interpretation of geometrically-weighted degree
  estimates in exponential random graph models.
\newblock {\em J Open Source Softw}, 1(3):36, 2016.

\bibitem{stivala20d}
Alex Stivala.
\newblock Reply to ``{C}omment on geodesic cycle length distributions in
  delusional and other social networks''.
\newblock {\em J Soc Struct}, 21(1):94--106, 2020.

\bibitem{ergm.tapered}
Mark~S. Handcock, Pavel~N. Krivitsky, and Ian Fellows.
\newblock {\em ergm.tapered: Tapered Exponential-Family Models for Networks},
  2022.
\newblock R package version 1.1-0.
  \url{https://github.com/statnet/ergm.tapered}.

\bibitem{stivala23_bionet}
Alex Stivala.
\newblock New network models facilitate analysis of biological networks.
\newblock {\em arXiv preprint arXiv:2312.06047v1}, 2023.

\bibitem{lolog}
Ian~E. Fellows.
\newblock {\em lolog: Latent Order Logistic Graph Models}, 2023.
\newblock R package version 1.3.1.
  \url{https://CRAN.R-project.org/package=lolog}.

\bibitem{agneessens04}
Filip Agneessens, Henk Roose, and Hans Waege.
\newblock Choices of theatre events: p* models for affiliation networks with
  attributes.
\newblock {\em Metod Zv}, 1(2):419--439, 2004.

\bibitem{berardo14}
Ramiro Berardo.
\newblock Bridging and bonding capital in two-mode collaboration networks.
\newblock {\em Policy Stud J}, 42(2):197--225, 2014.

\bibitem{bi21}
Youyi Bi, Yunjian Qiu, Zhenghui Sha, Mingxian Wang, Yan Fu, Noshir Contractor,
  and Wei Chen.
\newblock Modeling multi-year customers’ considerations and choices in
  {China}’s auto market using two-stage bipartite network analysis.
\newblock {\em Netw Spat Econ}, 21(2):365--385, 2021.

\bibitem{bond12}
Matthew Bond.
\newblock The bases of elite social behaviour: Patterns of club affiliation
  among members of the {House of Lords}.
\newblock {\em Sociology}, 46(4):613--632, 2012.

\bibitem{brandenberger15}
Laurence Brandenberger, Isabelle Schl{\"a}pfer, Philip Leifeld, and Manuel
  Fischer.
\newblock Interrelated issues and overlapping policy sectors: Swiss water
  politics.
\newblock \url{http://nbn-resolving.de/urn:nbn:de:bsz:352-0-294740}, 2015.

\bibitem{conaldi13}
Guido Conaldi and Alessandro Lomi.
\newblock The dual network structure of organizational problem solving: A case
  study on open source software development.
\newblock {\em Soc Netw}, 35(2):237--250, 2013.

\bibitem{desisto22}
Camille DeSisto and James~Paul Herrera.
\newblock Drivers and consequences of structure in plant--lemur ecological
  networks.
\newblock {\em J Anim Ecol}, 91(10):2010--2022, 2022.

\bibitem{duxbury18}
Scott~W. Duxbury and Dana~L. Haynie.
\newblock Building them up, breaking them down: Topology, vendor selection
  patterns, and a digital drug market’s robustness to disruption.
\newblock {\em Soc Netw}, 52:238--250, 2018.

\bibitem{duxbury18c}
Scott~W Duxbury and Dana~L Haynie.
\newblock The network structure of opioid distribution on a darknet
  cryptomarket.
\newblock {\em J Quant Criminol}, 34(4):921--941, 2018.

\bibitem{faust02b}
Katherine Faust, Karin~E Willert, David~D Rowlee, and John Skvoretz.
\newblock Scaling and statistical models for affiliation networks: patterns of
  participation among {Soviet} politicians during the {Brezhnev} era.
\newblock {\em Soc Netw}, 24(3):231--259, 2002.

\bibitem{fried22}
Harrison~S Fried, Matthew Hamilton, and Ramiro Berardo.
\newblock Closing integrative gaps in complex environmental governance systems.
\newblock {\em Ecol Soc}, 27(1):15, 2022.

\bibitem{fritz23}
Cornelius Fritz, Giacomo De~Nicola, Sevag Kevork, Dietmar Harhoff, and Göran
  Kauermann.
\newblock {Modelling the large and dynamically growing bipartite network of
  {German} patents and inventors}.
\newblock {\em J R Stat Soc Ser A Stat Soc}, 186(3):557--576, 03 2023.

\bibitem{fuzessy22}
Lisieux Fuzessy, Gisela Sobral, Daiane Carreira, Débora~Cristina Rother,
  Gedimar Barbosa, Mariana Landis, Mauro Galetti, Tad Dallas, Vinícius
  Cardoso~Cláudio, Laurence Culot, and Pedro Jordano.
\newblock Functional roles of frugivores and plants shape hyper-diverse
  mutualistic interactions under two antagonistic conservation scenarios.
\newblock {\em Biotropica}, 54(2):444--454, 2022.

\bibitem{gallemore15}
Caleb Gallemore, Monica {Di Gregorio}, Moira Moeliono, Maria Brockhaus, and
  Rut~Dini {Prasti H.}
\newblock Transaction costs, power, and multi-level forest governance in
  {Indonesia}.
\newblock {\em Ecol Econ}, 114:168--179, 2015.

\bibitem{gondal11}
Neha Gondal.
\newblock The local and global structure of knowledge production in an emergent
  research field: An exponential random graph analysis.
\newblock {\em Soc Netw}, 33(1):20--30, 2011.

\bibitem{gondal18}
Neha Gondal.
\newblock Duality of departmental specializations and {PhD} exchange: A
  {Weberian} analysis of status in interaction using multilevel exponential
  random graph models ({mERGM}).
\newblock {\em Soc Netw}, 55:202--212, 2018.

\bibitem{hamilton18}
Matthew Hamilton, Mark Lubell, and Emilinah Namaganda.
\newblock Cross-level linkages in an ecology of climate change adaptation
  policy games.
\newblock {\em Ecol Soc}, 23(2):36, 2018.

\bibitem{hazir12}
Cilem Hazir and Corinne Autant-Bernard.
\newblock Using affiliation networks to study the determinants of multilateral
  research cooperation: some empirical evidence from eu framework programs in
  biotechnology.
\newblock Working Paper 1212, GATE, 2012.
\newblock \url{https://doi.org/10.2139/ssrn.2060275}.

\bibitem{heaney18_preprint}
Michael~T. Heaney and Philip Leifeld.
\newblock Contributions by interest groups to lobbying coalitions.
\newblock {\em J Polit}, 80(2):494--509, 2018.

\bibitem{huang09}
Yun Huang, Mengxiao Zhu, Jing Wang, Nishith Pathak, Cuihua Shen, Brian Keegan,
  Dmitri Williams, and Noshir Contractor.
\newblock The formation of task-oriented groups: Exploring combat activities in
  online games.
\newblock In {\em 2009 International Conference on Computational Science and
  Engineering}, volume~4, pages 122--127, 2009.

\bibitem{jasny12}
Lorien Jasny.
\newblock Baseline models for two-mode social network data.
\newblock {\em Policy Stud J}, 40(3):458--491, 2012.

\bibitem{jasny15}
Lorien Jasny and Mark Lubell.
\newblock Two-mode brokerage in policy networks.
\newblock {\em Soc Netw}, 41:36--47, 2015.

\bibitem{keegan12}
Brian Keegan, Darren Gergle, and Noshir Contractor.
\newblock Do editors or articles drive collaboration? multilevel statistical
  network analysis of {Wikipedia} coauthorship.
\newblock In {\em Proceedings of the ACM 2012 Conference on Computer Supported
  Cooperative Work}, CSCW '12, page 427–436, New York, NY, USA, 2012.
  Association for Computing Machinery.

\bibitem{khalilzadeh18}
Jalayer Khalilzadeh.
\newblock Demonstration of exponential random graph models in tourism studies:
  Is tourism a means of global peace or the bottom line?
\newblock {\em Ann Tour Res}, 69:31--41, 2018.

\bibitem{lai17}
Chih-Hui Lai, Chen-Chao Tao, and Yu-Chung Cheng.
\newblock Modeling resource network relationships between response
  organizations and affected neighborhoods after a technological disaster.
\newblock {\em Voluntas}, 28(5):2145--2175, 2017.

\bibitem{leifeld14_preprint}
Philip Leifeld and Thomas Malang.
\newblock National parliamentary coordination after {Lisbon}: A network
  approach.
\newblock Paper prepared for the 1st European Conference on Social Networks
  (EUSN), Barcelona, Spain, July 2014.

\bibitem{li21b}
Qingchun Li and Ali Mostafavi.
\newblock Local interactions and homophily effects in actor collaboration
  networks for urban resilience governance.
\newblock {\em Appl Netw Sci}, 6(1):89, 2021.

\bibitem{lu22}
Yingjie Lu and Qian Wang.
\newblock Doctors’ preferences in the selection of patients in online medical
  consultations: an empirical study with doctor--patient consultation data.
\newblock {\em Healthcare}, 10(8):1435, 2022.

\bibitem{lubell14}
Mark Lubell, Garry Robins, and Peng Wang.
\newblock Network structure and institutional complexity in an ecology of water
  management games.
\newblock {\em Ecol Soc}, 19(4):23, 2014.

\bibitem{margolin12}
Drew Margolin, K~Ognyanoya, Meikuan Huang, Yun Huang, and Noshir Contractor.
\newblock Team formation and performance on {nanoHub}: a network selection
  challenge in scientific communities.
\newblock In Bal\'{a}zs Vedres and Marco Scotti, editors, {\em Networks in
  Social Policy Problems}, chapter~5, page 80–100. Cambridge University
  Press, 2012.

\bibitem{mauldin21}
Rebecca~L Mauldin, Carin Wong, Jason Fernandez, and Kayo Fujimoto.
\newblock Network modeling of assisted living facility residents’ attendance
  at programmed group activities: Proximity and social contextual correlates of
  attendance.
\newblock {\em Gerontologist}, 61(5):703--713, 2021.

\bibitem{mcallister14}
Ryan~RJ McAllister, Rod McCrea, and Mark~N Lubell.
\newblock Policy networks, stakeholder interactions and climate adaptation in
  the region of {South East Queensland, Australia}.
\newblock {\em Reg Environ Change}, 14(2):527--539, 2014.

\bibitem{mcallister15}
Ryan~RJ McAllister, Catherine~J Robinson, Kirsten Maclean, Angela~M Guerrero,
  Kerry Collins, Bruce~M Taylor, and Paul~J De~Barro.
\newblock From local to central: a network analysis of who manages plant pest
  and disease outbreaks across scales.
\newblock {\em Ecol Soc}, 20(1):67, 2015.

\bibitem{mcallister15b}
Ryan~RJ McAllister, Bruce~M Taylor, and Ben~P Harman.
\newblock Partnership networks for urban development: how structure is shaped
  by risk.
\newblock {\em Policy Stud J}, 43(3):379--398, 2015.

\bibitem{mcallister17}
Ryan~RJ McAllister, Catherine~J Robinson, Alinta Brown, Kirsten Maclean, Suzy
  Perry, and Shuang Liu.
\newblock Balancing collaboration with coordination: contesting eradication in
  the {Australian} plant pest and disease biosecurity system.
\newblock {\em Int J Commons}, 11(1):330--354, 2017.

\bibitem{metz18}
Florence Metz, Philip Leifeld, and Karin Ingold.
\newblock Interdependent policy instrument preferences: a two-mode network
  approach.
\newblock {\em J Public Policy}, 39(4):609--636, 2019.

\bibitem{niekamp13}
Anne-Marie Niekamp, Liesbeth~AG Mercken, Christian~JPA Hoebe, and Nicole~HTM
  Dukers-Muijrers.
\newblock A sexual affiliation network of swingers, heterosexuals practicing
  risk behaviours that potentiate the spread of sexually transmitted
  infections: a two-mode approach.
\newblock {\em Soc Netw}, 35(2):223--236, 2013.

\bibitem{nita16}
Andreea Nita, Laurentiu Rozylowicz, Steluta Manolache, Cristiana~Maria
  Cioc{\u{a}}nea, Iulia~Viorica Miu, and Viorel~Dan Popescu.
\newblock Collaboration networks in applied conservation projects across
  {Europe}.
\newblock {\em PLoS One}, 11(10):e0164503, 2016.

\bibitem{norbutas18}
Lukas Norbutas.
\newblock Offline constraints in online drug marketplaces: An exploratory
  analysis of a cryptomarket trade network.
\newblock {\em Int J Drug Policy}, 56:92--100, 2018.

\bibitem{park22}
Sejung Park and Rong Wang.
\newblock Assessing the capability of government information intervention and
  socioeconomic factors of information sharing during the {COVID}-19 pandemic:
  a cross-country study using big data analytics.
\newblock {\em Behav Sci}, 12(6):190, 2022.

\bibitem{ren23}
Hang Ren, Lu~Zhang, Travis~A Whetsell, and N~Emel Ganapati.
\newblock Analyzing multisector stakeholder collaboration and engagement in
  housing resilience planning in greater {Miami} and the beaches through social
  network analysis.
\newblock {\em Nat Hazards Rev}, 24(1):04022036, 2023.

\bibitem{rocha15}
Juan~Carlos Rocha, Garry~D Peterson, and Reinette Biggs.
\newblock Regime shifts in the {Anthropocene}: drivers, risks, and resilience.
\newblock {\em PLoS One}, 10(8):e0134639, 2015.

\bibitem{scott17}
Tyler~A Scott and Craig~W Thomas.
\newblock Winners and losers in the ecology of games: Network position,
  connectivity, and the benefits of collaborative governance regimes.
\newblock {\em J Public Adm Res Theory}, 27(4):647--660, 2017.

\bibitem{sha19}
Zhenghui Sha, Youyi Bi, Mingxian Wang, Amanda Stathopoulos, Noshir Contractor,
  Yan Fu, and Wei Chen.
\newblock Comparing utility-based and network-based approaches in modeling
  customer preferences for engineering design.
\newblock In {\em Proceedings of the Design Society: International Conference
  on Engineering Design}, volume~1, pages 3831--3840, 2019.

\bibitem{sha19b}
Zhenghui Sha, Ashish~M Chaudhari, and Jitesh~H Panchal.
\newblock Modeling participation behaviors in design crowdsourcing using a
  bipartite network-based approach.
\newblock {\em J Comput Inf Sci Eng}, 19(3):031010, 2019.

\bibitem{shin22}
Seongmin Shin, Mi~Sun Park, Hansol Lee, and Himlal Baral.
\newblock The structure and pattern of global partnerships in the {REDD}+
  mechanism.
\newblock {\em For Policy Econ}, 135:102640, 2022.

\bibitem{shjarback18}
John~A Shjarback and Jacob~TN Young.
\newblock The “tough on crime” competition: A network approach to
  understanding the social mechanisms leading to federal crime control
  legislation in the {United States} from 1973--2014.
\newblock {\em Am J Crim Just}, 43(2):197--221, 2018.

\bibitem{stephens16}
Bryan Stephens, Wenhong Chen, and John~Sibley Butler.
\newblock Bubbling up the good ideas: A two-mode network analysis of an
  intra-organizational idea challenge.
\newblock {\em J Comput Mediat Commun}, 21(3):210--229, 2016.

\bibitem{wang20b}
Rong Wang.
\newblock Marginality and team building in collaborative crowdsourcing.
\newblock {\em Online Inf Rev}, 44(4):827--846, 2020.

\bibitem{wonka19}
Arndt Wonka and Sebastian Haunss.
\newblock Cooperation in networks: political parties and interest groups in
  {EU} policy-making in {Germany}.
\newblock {\em Eur Union Polit}, 21(1):130--151, 2020.

\bibitem{zhu13}
Mengxiao Zhu, Yun Huang, and Noshir~S Contractor.
\newblock Motivations for self-assembling into project teams.
\newblock {\em Soc Netw}, 35(2):251--264, 2013.

\end{thebibliography}

\end{document}